\documentclass[conference]{IEEEtran}

\usepackage{graphicx}
\usepackage{balance}  
\usepackage{times,amsmath,amsthm,epsfig,cite,soul,enumitem}
\usepackage{caption,algpseudocode,algorithm,setspace}
\usepackage{subcaption}
\usepackage{adjustbox}
\usepackage[normalem]{ulem}
\usepackage[usenames, dvipsnames]{color}
\usepackage{epstopdf}
\usepackage{color,soul}

\makeatletter
\renewcommand{\ALG@beginalgorithmic}{\footnotesize}
\makeatother

\def\NoNumber#1{{\def\alglinenumber##1{}\State #1}\addtocounter{ALG@line}{-1}}

\ifCLASSINFOpdf
\else
\fi
\begin{document}
%
\title{Fault Tolerant Frequent Pattern Mining}


\author{\IEEEauthorblockN{Sameh Shohdy\IEEEauthorrefmark{1},
Abhinav Vishnu\IEEEauthorrefmark{2} and
Gagan Agrawal\IEEEauthorrefmark{1}}
\IEEEauthorblockA{\IEEEauthorrefmark{1}Department of Computer Science
and Engineering\\
The Ohio State University,
Columbus, OH 43210, \{ahmedsa, agarwal\}@cse.ohio-state.edu}
\IEEEauthorblockA{\IEEEauthorrefmark{2}High Performance Computing
Group\\
Pacific Northwest National Laboratory,
Richland, WA 99352, abhinav.vishnu@pnnl.gov}}


%


\maketitle

\baselineskip=0.90\normalbaselineskip
\begin{abstract}

FP-Growth algorithm is a Frequent Pattern Mining (FPM) algorithm  that
has been extensively used to study correlations and patterns in large
scale datasets. While several researchers have designed distributed
memory FP-Growth algorithms, it is pivotal to consider fault tolerant
FP-Growth, which can address the increasing fault rates in large scale
systems.
In this work, we propose a novel {
parallel, algorithm-level fault-tolerant} FP-Growth algorithm. We
leverage algorithmic properties and MPI advanced features to guarantee an
$O(1)$ space complexity, achieved by using the dataset memory space
itself for checkpointing. We also propose a recovery algorithm  that can
use in-memory  and disk-based checkpointing,  though in many cases the
recovery can be completed without any disk access, and incurring no
memory overhead for checkpointing. We evaluate our FT algorithm on a
large scale InfiniBand cluster with several large datasets using up to
2K cores. Our evaluation
demonstrates excellent efficiency for checkpointing and recovery in
comparison to the disk-based approach. We have also observed {\bf 20x}
average speed-up in comparison to Spark, establishing that a well
designed
algorithm can easily outperform a  solution based on a general
fault-tolerant  programming model.

\end{abstract}

\IEEEpeerreviewmaketitle

\section{Introduction}
\label{sec:intro}
Machine Learning and Data Mining (MLDM) algorithms are becoming
ubiquitous in analysing large volume of data produced  in science areas
(instrument and simulation data) as well as other areas such as  social
networks and  financial transactions.  Frequent Pattern Mining (FPM) is
an important MLDM algorithm, which is used for finding attributes that
frequently occur together. Due to its high applicability, several FPM
algorithms have been proposed in the literature such as
Apriori~\cite{buehrer2007toward}, Eclat~\cite{zaki1997parallel},
FP-Growth~\cite{han2000mining}, and GenMax~\cite{gouda2005genmax}.
However, FP-Growth has become  extremely popular due to its relatively
small space and time complexity requirements. 

To address increasing data volumes, several researchers have proposed
large scale distributed memory FP-Growth
algorithms~\cite{fpgrowth:cluster, fang:damon09,
li:recsys08,buehrer:ppopp07,vishnu:cluster15}.  One of the challenges
that arise with execution  on large-scale parallel systems is the
increased likelihood (and frequency) of faults.  Large scale systems
frequently suffer from faults of several types in many components
~\cite{sridharan:sc12, sridharan:sc13,
schroeder:dsn06,vandam:jctc13,bronevetsky:ics08,shantharam:ics11}.

Driven by these trends, several recent programming models such as
Hadoop, Spark~\cite{spark},  and MillWheel~\cite{millwheel} have
considered  fault tolerance to be  one of the most  important   design
consideration.  Hadoop achieves fault tolerance by using multiple
replicas of the data structures in permanent storage --- possibly
resulting in a significant amount of I/O in the critical path. Spark
addresses this limitation by using Resilient Distributed Datasets
(RDDs), such that in-memory replication can be used for fault tolerance.
However, for very large datasets, in-memory replication is infeasible.
In several cases, Spark considers disk as the backend for checkpointing
--- which can again significantly slow-down the computation and increase
data movement.  Similarly, MillWheel is used for fault tolerant stream
processing and uses the disk as the backend for checkpointing.
Naturally, an advantage of using fault tolerant programming model is the
fact that checkpointing and recovery is automated.  However, the
performance penalty of a fault tolerant programming model (due to
disk-based checkpointing) or space overhead (due to in-memory
checkpointing) is unattractive for scaling several MLDM algorithms at
large volume and computing scale.

In the context of general-purpose  programming systems, recently
proposed  methods such as Scalable Checkpoint Restart (SCR)
~\cite{moody2010design} are able to provide in-memory checkpointing for
multi-level hierarchical file systems using non-blocking methods. SCR
also allows using spare main memory for in-memory checkpointing.
Similarly, other researchers have proposed programming model/runtime
extensions to Charm++, and X10 for supporting fault tolerance.  While
these approaches provide non-blocking checkpointing,  the overall memory
requirements increase, since the implementations need to use spare
memory for checkpointing. This can very well make the approach
infeasible,  especially with weak scaling executions, where spare memory
is scarce.

\begin{figure}[!htb]
\centerline{\psfig{figure=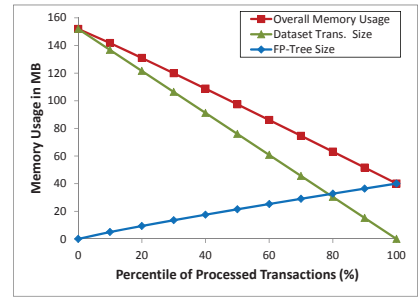,width=55mm} }
\caption{\small{Pattern of Memory Requirements of FP-Tree and Dataset during FP-Tree build phase. As more transactions are processed, lesser
memory is required for dataset --- which can be used for checkpointing}}
\label{fig:motivation}
\end{figure}

In this paper, we present an in-depth study  of FP-Growth algorithm
for fault tolerance. Considering its two-pass properties (impact shown
in Figure~\ref{fig:motivation}), we propose a
novel algorithm, which requires $O(1)$ space complexity for saving
critical data structures, i.e., FP-Tree, in memory of other computing
nodes. The proposed algorithm incrementally leverages the memory
allocated for the default algorithm for checkpointing FP-Trees -- and possibly partial
replica of transactions from other computing nodes -- ensuring an $O(1)$
space overhead of our proposed algorithms. To further minimize time
overhead for checkpointing, our solution not only leverages non-blocking
properties, but use MPI-Remote Memory Access ( MPI-RMA) in
addition to minimize any involvement of remote process for checkpointing.
By using MPI-RMA and contiguous data structures for implementing our
proposed algorithms, we are able to leverage Remote Direct Memory Access
(RDMA) effectively. We believe that our proposed extensions may be
included with existing solutions such as SCR, where a class of
algorithms may re-use already allocated memory for checkpointing and
recovery.

\subsection{Contributions}
Specifically, we make the following contributions in the paper:

\begin{itemize}
		\item We propose an $O(1)$ in-memory checkpointing based
				FP-Growth algorithm for large scale systems.  The
				proposed algorithm leverages overlapping communication
				with FP-Tree build phase --- such that the overhead of
				checkpointing is minimized.
\item  We propose three different fault tolerance parallel FP-Growth mechanisms:  a default Disk-based Fault
		tolerant FP-Growth  (DFT), Synchronous Memory-based Fault
		tolerant FP-Growth (SMFT), and an Asynchronous Memory-based
		Fault tolerant FP-Growth (AMFT).
\item We study the limitations of existing programming models (Hadoop
		MapReduce, Spark and MillWheel) and implement our algorithms
		using Message Passing Interface (MPI)~\cite{mpi1,mpi2}.
		Specifically, we use MPI-RMA mechanism to checkpoint critical
		data structures of FP-Growth asynchronously. With recent
		developments in MPI-RMA Fault tolerance~\cite{besta:hpdc14}, it
		is possible to use MPI for handling faults, while providing
		native performance.
\item We perform an in-depth evaluation of our proposed approaches
		using up to 200M transactions and 2048 cores. 
 Using 100M transactions on 2048 cores, the
		checkpointing overhead is $\approx$ 5\%, while the recovery cost
		for multiple failures is independent of the number of processes.
\item We also show the effectiveness of our fault-tolerant FP-Growth
		implementation -- implementations outperforms Spark
		implementations of the same algorithm by providing {\bf 20x}
		average speed-up. 
\end{itemize}


\section{Preliminaries}
\label{sec:background}

\subsection{Frequent Pattern Mining}

Frequent Pattern Mining (FPM) algorithms find items that  frequently
occur together within transactions of a database. An item or  itemset is defined as
frequent if its frequency is higher than  a {\em user-defined threshold}.
Several FPM algorithms have been proposed in the literature including
Apriori, Eclat, GenMax and FP-Growth. The FP-Growth algorithm is  very
popular since it requires only two passes on the dataset, does not
involve  candidate generation (unlike Apriori) and provides a compressed
representation of the frequent items using a {\em Frequent Pattern
(FP)-Tree}. We specifically focus on designing  parallel fault-tolerant
versions of the FP-Growth algorithm, due to its attractive properties.

During the first pass, FP-Growth algorithm finds items  that  occur
frequently. In the second pass, it creates an FP-Tree, which is a
modified {\em Trie}. The first pass requires a simple scan through the
given dataset to find all single frequent items.  {\bf FP-Tree creation
step (the second pass) is the most time consuming part of the overall
calculation}~\cite{vishnu:cluster15}. Hence, we focus on fault tolerant
FP-Tree creation step of the algorithm, since longer execution time also implies higher fault
probability.

\subsection{Faults}

Large scale systems suffer from several fault types --- permanent,
transient,  and intermittent. A permanent fault typically requires a
device (such as a compute node) to be replaced. We consider fault
tolerance for permanent process faults in this paper.
We assume a fail-stop fault model --- once a process is perceived as {\em dead/faulty}, it is
presumed unavailable for the rest of the computation.

Since permanent node faults are commonplace in large scale systems,
several researchers have proposed techniques for addressing these
faults. Typically, {\em
checkpoint-restart}~\cite{saini2014affinity,bronevetsky2003automated}
based methodologies are used. Application-independent methods checkpoint
the entire application space on a permanent disk --- however, they have
been shown to scale only on small size
systems~\cite{bronevetsky2003automated}. Application-dependent methods
--- also known as Algorithm Based Fault Tolerance
(ABFT)~\cite{chen2011algorithm,lisboa2008algorithm,shohdy2016parallel,abdulah2016addressing}
methods reduce this overhead by selectively checkpointing important data
structures periodically.  However, depending up on the application
characteristics, checkpointing of critical data structures may still
require disk access.

\subsection{Fault Tolerant Programming Models}

Recently, there has been a surge of large scale and fault tolerant
functional programming models  such as Hadoop, Spark, and MillWheel.
Functional programming, in turn,  uses  the concept of single
assignment, where every mutation of a variable is recorded, saved (on a
permanent storage/memory of another node),  and replayed when a fault
occurs. 

Now, let us examine the implication of such a  framework for an
algorithm like FP-Tree. Every change or mutation needs to be  recorded locally, and such records can be eventually saved to permanent storage. In many cases, the step of saving a new version
of the FP-Tree on the disk is carried-out    at the end of the Reduce
phase (of the MapReduce implementation). For a two-phase algorithm such
as FP-Tree, where most of the time is spent on the second phase,  no
advantage is achieved.  Another possible implementation may choose to
divide the overall computation into multiple MapReduce steps. The
checkpointing can be executed at the end of each Reduce phase. However,
now  the overall execution time will increase, since saving a new
version      will either involve writing to a disk (expensive) or
neighbor's memory. Since the reduce phase is a blocking phase, the
application will observe a significant  overhead of checkpointing, which
will   degrade the overall performance.  Naturally, a scalable algorithm
should harness best possible performance by using native execution,
while minimizing the cost of checkpointing, by using {\em non-blocking}
methods.


Now,  in  examining  an alternate programming model, we  consider the Message
Passing Interface (MPI)~\cite{mpi1, mpi2}, which has been readily
available and widely used on  supercomputers and clusters, and beginning to find its place on cloud computing systems.
 While MPI has been frequently criticized for lack of fault
tolerance support, recent literature and implementations indicate that fault
tolerance is addressed well for permanent process
faults~\cite{besta:hpdc14}. More importantly, recently introduced MPI
One-sided - MPI one-sided communication (also known as MPI-Remote Memory Access
 (MPI-RMA))~\cite{mpi1,mpi2}- primitives provide necessary tools for overlapping
communication with computation. With this observation, we focus on using  MPI for designing fault
tolerant FP-Growth algorithm in this paper.

\section{Parallel Baseline Algorithm}
\label{sec:sspace}
Algorithm~\ref{alg:PFPTREE} shows the key steps of the parallel
FP-Growth
algorithm, which we have used as the baseline for designing fault
tolerant FP-Growth algorithms.

A brief explanation of the steps is presented here: The first step is
to distribute the input database transactions among  $|P|$ processes
(Line 3) (Each process is a worker, which is involved in computing
its
{\em local} FP-Tree).  Each process ($p_i$) scans the local transactions
and records the frequency of each item (Line 4). To collect the global frequency,
an all-to-all reduction (by \texttt{MPI\_Allreduce}) is used
(incurring
$\log(|P|)$ time complexity) (Line 5). After all-to-all reduction, the items with
frequency greater than support threshold are saved, and other items are
discarded.
Then, each $p_i$ generates a local FP-Tree ($L.Tree$) using its local
transactions, which have at least one frequent item (Line 6).  Later,
each $p_i$ merges its local FP-Tree with the FP-Trees from other
processes to produce a global FP-Tree ($G.Tree$) by using a ring
communication algorithm~\cite{vishnu:cluster15} (Line 7).
Finally, frequent itemsets
($FreqItemSet$) are produced using the output global FP-Tree (Line 8).

\begin{algorithm} [htp]
\setstretch{1.0}
  \caption{\bf : Parallel FP-Growth Algorithm }
  \label{alg:PFPTREE}
\begin{algorithmic}[1]
\State {\bf Input:} Set of transactions $S$, Support threshold $\theta$
\State {\bf Output:} Set of frequent itemsets
\State L.Trans $\leftarrow$    getLocalTrans($S$)
\State L.FreqList $\leftarrow$    findLocalFreqItems($L.Trans$, $\theta$)
\State G.FreqList $\leftarrow$    Reduce Local Freq items through all processes
\State L.Tree $\leftarrow$    generateLocalFPTree($L.Trans$, $G.Freq.List$)
\State G.Tree $\leftarrow$    generateGlobalFPTree($L.Tree$)
\State FreqItemSet $\leftarrow$    miningGFPTree($G.Tree$)
\end{algorithmic}
\end{algorithm}

Further,  we summarize the symbols we have used to model the time and space complexity of the proposed fault tolerant algorithms
in Table~\ref{table:nonlin}.

\tabcolsep=0.11cm
\begin{table}[ht]
\small
\caption{Symbols used for Time-Space Complexity Modeling} 
\centering 
\begin{tabular}{|c|c| } 
\hline\hline 
Name & Symbol \\ [1 ex] 
\hline 
Process Set & $P = {\{p_{0} \cdots p_{|P| - 1} \}}$  \\ 
Transaction Database & $T = \{t_{0} \cdots t_{|T| - 1}\}$ \\
Average Local Transaction Size & $t_{avg}$ \\
Minimum Support Threshold & $\theta$ \\
Local FP-Tree Set & $S = \{ s_{0} \cdots s_{|P| - 1}\} $ \\
Average Local FP-Tree size & $s_{avg} $ \\
Average time to merge two local FP-Trees & $m$ \\
Number of Checkpoints & $C$\\
Disk Access Bandwidth & $l$ \\
Network Bandwidth & $b$ \\ [1ex] 
\hline 
\end{tabular}
\label{table:nonlin} 
\end{table}

\section{Proposed FP-Growth Fault Tolerant Algorithms}
\label{sec:design}
In this section, we present several approaches for designing fault
tolerant FP-Growth algorithm. Our baseline algorithm uses the disk as the
safe storage for saving intermediate FP-Trees, whereas the
optimized algorithms  use the memory originally allocated to the
database transactions for checkpointing intermediate FP-Trees and
transactions of other processes (with a high overlap  of communication
with computation achieved using MPI-RMA methodology).

To design a fault tolerant FP-Growth algorithm, there are several design
choices. Since we consider fail-stop model, it is important to
understand the design choices between {\em re-spawning} a new set of
processes on a spare node versus {\em continued-execution} with existing
processes and nodes. We use continued-execution, primarily because for
most systems, it is intricate to re-spawn, attach the processes/node to
the existing set of processes, and continue recovery.  Instead,
continued-execution provides a simple mechanism to conduct recovery,
without significant dependence on external software.

\subsection{Disk-based Fault Tolerant (DFT) FP-Gro\-wth}

The Disk-based Fault Tolerant (DFT) algorithm is
the  baseline   for other approaches presented in this
paper.

\noindent
{\bf Checkpointing Algorithm and Complexity:}
In the FP-Growth algorithm, there are two critical data
structures that  are needed during the recovery process --- database transactions themselves
and intermediate FP-Trees generated by the processes.
Under the DFT approach, the intermediate FP-Trees generated by each
process are periodically saved on disk. For many supercomputers, the
disks are located remotely, such as a remote storage. In other cases,
locally available SSDs can be used as well. The database transactions
are already resident on the disk. Hence, it is not necessary to
checkpoint the database transactions.

Let us consider an equal
distribution of database transactions to processes ($|T|/|P|$
transactions are available on each process). Let $C$ be the number of
checkpoints, which are executed by the application. The number of
checkpoints are derived as a function of $|T|$, and $|P|$, such that the
cost of checkpointing can be amortized over the FP-Tree creation phase.
The DFT algorithm also needs to save metadata file associated with
FP-Tree, which may be used during recovery. The space complexity of the
metadata file is negligible, since only a few integers need to be
saved.

Let $s_{avg}$ represent the average size of an FP-Tree generated by each
process (calculated as $\frac{\sum_{i=0}^{|P| - 1} s_i}{ |P|}$).
The time complexity for checkpointing intermediate FP-Trees is
$O(\frac{C\cdot s_{avg}}{l})$. However, the actual time to checkpoint
can escalate due to the contention from multiple processes writing the
checkpoint file simultaneously. The space complexity incurred by each
process is $O({C\cdot s_{avg}})$, which can be reduced further by
recycling existing checkpoints.

\noindent
{\bf Recovery Algorithm and Complexity:}
In the DFT approach, the recovery is initiated by the master ($p_m$) (In
our implementation we use the default process --- process with the first
rank in MPI as the master). $p_m$ reads the metadata file associated
with the faulty process ($p_f$), which provides the necessary information
for conducting recovery. A recovery process ($p_r$) is selected, which
reads checkpointed FP-Tree of $p_f$ from the disk and merges the
checkpointed FP-Tree of $p_f$ with its FP-Tree, while $p_m$ reads dead
process transactions from disk, and re-distributes them among remaining
processes.

The time complexity of the recovery algorithm is a function of reading
the partial dataset and executing the recovery algorithm. In the
worst case, the entire transactions of the faulty process need to be
re-executed. Hence, the worst case time complexity is
$\frac{|T|}{|P|\cdot l}$ (reading the dataset) $ + \frac{|T|}{|P|
\cdot b}$ (re-distributing among process)  + $m$
(re-computation), where $m$ is the average cost of merging a transaction
in an existing FP-Tree (In the worst case, the FP-Tree is null, since all
transactions are re-executed).


\noindent
{\bf Implementation Details:}
{As mentioned earlier, each process saves a copy of local
FP-Tree in a safe storage. Thus, our implementation depends on
checkpointing local FP-Tree on disk --- $LFP_{Backup}$ file. This file
associated with another $metadata$ file describes the checkpointed
FP-Tree by storing a set of description values such as: checkpoint
timestamp and last processed transaction. Each process asynchronously
updates both files, during the
execution. In the case of failure, the recovery operation is
performed in two steps: The pre-determined recovery process $p_r$
process reads the last checkpointed FP-Tree of the faulty process $p_f$
from the disk and merges it with its local FP-Tree. At the same time,
the master process reads the metadata file of $p_f$ to decide the set of
transactions to be recovered from the disk. The master process recovers
unprocessed transactions and redistributes them to the remaining
processes.

\noindent
{\bf Advantages and Limitations of DFT:}
The proposed DFT algorithm is largely equivalent to designing a fault
tolerant FP-Growth algorithm using MapReduce programming models such as
Hadoop/Spark. However, an advantage is that it  can specifically take
advantage of native communication by using MPI, especially when high
performance interconnects are available.  Disk-based
approach makes DFT suffer from several limitations: These include
prohibitive I/O cost for checkpointing/recovering local FP-Trees and
recovering unprocessed transactions, and centralized bottleneck of the
master process in the case of failure to re-read unprocessed
transactions from the disk.

\subsection{Synchronous Memory-based Fault Tolerant (SMFT) FP-Growth}
As discussed above, the primary limitation of the DFT approach is that
it uses disk-based checkpointing and recovery,   which is prohibitive
for scaling the FP-Growth algorithm. Hence, it is important to consider
memory based fault tolerant FP-Growth algorithm.

Since available memory size is relatively small in comparison to the
disk size, it is also unattractive to incur additional space complexity
for in-memory checkpointing of FP-Trees and database transactions from
other processes.  SMFT involves checkpointing method where the overall
space complexity   of the algorithm remains constant. Additionally, we
overlap the checkpointing of FP-Trees and database transactions by using
non-blocking primitives provided by the MPI one-sided model. We present
the checkpointing, and recovery methods with their time-space complexity
analysis in the ensuing  sections.

\begin{figure*}
\centerline{\psfig{figure=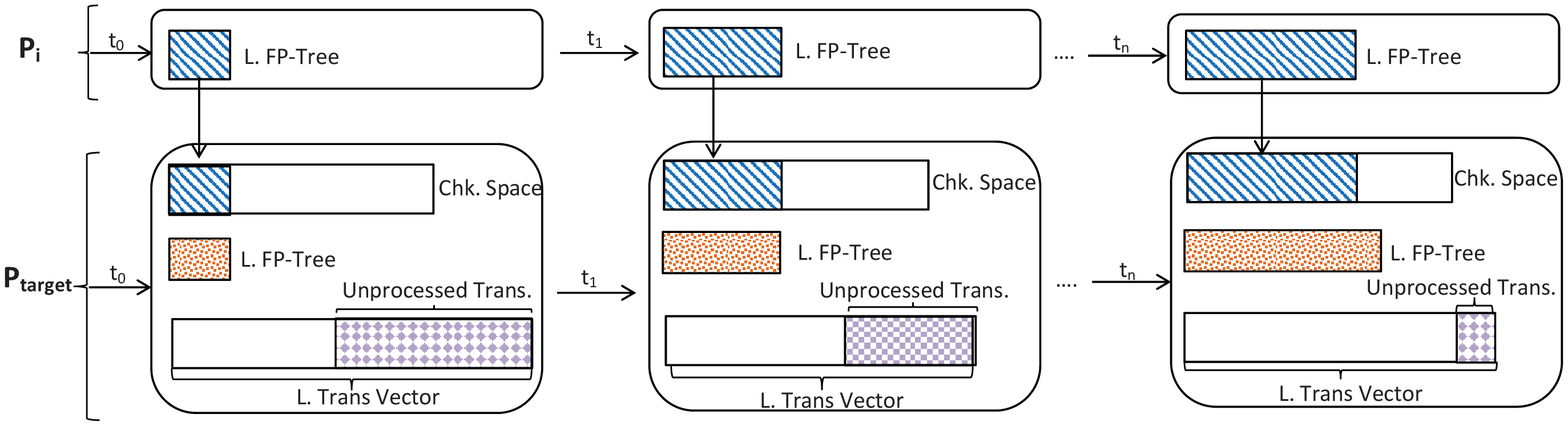,width=130mm} }
\caption{SMFT FP-Tree Checkpointing Operation Overview}
\label{fig:SMFT}
\end{figure*}
\noindent
{\bf Checkpointing Algorithm:}
The premise of constant space complexity is based on the two-pass
properties of the FP-Growth algorithm. During the FP-Tree creation
phase, once a database transaction is processed, the memory occupied by
the transaction can be used for checkpointing. We leverage this property
of the algorithm to checkpoint the FP-Trees and database transactions.
Specifically, once a transaction is processed, we reclaim the memory
consumed by the transaction and allocate a separate {\em window} of
memory, which can be used by other processes for checkpointing their
FP-Trees and database transactions. With this technique, the overall
space complexity of the algorithm is $O(1)$.

Besides optimal space complexity, the objective of SMFT algorithm is to minimize
the time complexity of checkpointing both the FP-Trees and database
transactions. Considering $C$ as the number of checkpoints, under a
naive algorithm, each process can checkpoint its existing FP-Tree to
another process at every $\frac{|T|}{|P|\cdot C}$ steps. Since  the time
overhead of checkpointing is non-negligible, as this step   {\em blocks}
for the communication to complete before continuing to process remaining
transactions,  at every checkpointing step --- with increasing FP-Tree
size --- the overhead of blocking increases. Hence, it is important to
consider non-blocking methods of checkpointing, such that communication
cost of checkpointing can be overlapped with computation.

SMFT algorithm uses MPI one-sided non-blocking methods for checkpointing. Specifically, as the database transactions are processed, a similar amount of memory is added to a {\em checkpoint window}. The algorithm uses dynamic allocation feature in MPI-RMA , \texttt{MPI\_Win\_create\_dynamic}, that allows incremental increase in the size of the checkpointing memory space during the execution. However, this dynamic allocation technique requires synchronization between both cooperated processes to perform each single checkpoint which adds more overhead to the checkpoint process. SMFT checkpoint overhead comes from different sources: waiting time till synchronization, communication ---which is negligible based on well known communication model LogGP}~\cite{alexandrov:95}---, and memory allocation and de-allocation cost.



Figure~\ref{fig:SMFT} shows an overview over the FP-Tree checkpointing operation in
SMFT approach. Assuming process $p_i$ needs to checkpoint on process
$p_{target}$ memory, each time period, i.e., $t_0$, $t_1$, ..., $t_n$,
process $p_{target}$ re-initiates a checkpoint space that can handle
process $p_i$ checkpointed local FP-Tree. In this case, process $p_i$
can remotely checkpoint its local FP-Tree to the new assigned location
without communicating with checkpoint process $p_{target}$.

\noindent
{\bf Recovery Algorithm:}
Assuming a process $p_{f}$ fails while executing the FP-Tree
phase. On fault recovery, the recovery process $p_r$ (in the simplistic
case, a neighbor such as $p_{f+1}$) merges checkpointed FP-Tree of $p_f$
stored on its memory to its local FP-Tree. If $p_{r}$ has also stored
part of the database transactions from $p_f$, it re-distributes these
transactions to other processes, which are still active in the
computation. The recovered transactions can be gathered from the memory
of $p_{r}$, if they were checkpointed by $p_{f}$ before failure. In the
case of disk recovery, lost transactions can be read from the disk using
two different ways. First, dataset transactions may be read from the
disk by using the master process and re-distributed evenly among the
remaining processes. However, in this case, disk access will be the most
expensive part of the overall recovery algorithm. So, we suggest using
all available processes to read samples of failed process ($\frac{n}{p}$) from the disk in parallel. With
this, each process will only access the disk to read $\frac{n}{P(P-1)}$
transactions. Further, since failed process $p_f$ held the data
checkpointed by process $p_{f-1}$ , process $p_{f-1}$ performs a critical
checkpoint on process $p_{rec}$ ---  in the simplest case, the processes can be assumed
to be connected in a virtual ring topology. Using this methodology, there is always
at least one replica of the FP-Tree of each process.

\noindent
{\bf Advantages and Limitations of SMFT:}
The primary advantage of SMFT is that it avoids reading/writing from the disk. Naturally, SMFT achieves native performance using MPI and is expected to incur low overhead for checkpointing with non-blocking MPI one-sided communication. The recovery algorithm uses memory to recover the database transactions, if possible. By distributing the transactions of a failed process to other active processes, the algorithm is able to minimize the recovery overhead. In the case of disk-based transactions recovery, SMFT uses all processes to read recovered transactions from the disk in parallel to avoid master process bottleneck.

SMFT approach has two main limitations. First, each two processes $p_i$ and $p_{target}$ need to synchronize in all checkpoints to share the address of checkpoint vector and the size of checkpointed FP-Tree or checkpointed transactions. Second, SMFT algorithm requires de-allocating existing space and allocating new space for checkpointing window. The overhead of synchronization, de-allocation and allocation are observed during FP-Tree creation phase. We address these two limitations in the AMFT approach, presented later.

\noindent
{\bf Implementation Details:}
In SMFT, each process $p_{target}$ allocates three memory vectors. These vectors are used to handle checkpoints from process $p_i$ namely: $FPT.chk_{target}$ vector to handle local FP-Tree of proceeding process $p_{i}$, $ Trans.chk_{target}$ vector to handle transactions checkpoint of $p_{i}$, and $metadata_{target}$ vector that includes a set of parameters to describe both checkpoint vectors. These vectors are allocated and exposed for read/update by each process using MPI-RMA primitives.

For in-memory checkpointing, SMFT requires that each process $p_i$ selects another process for checkpointing. While SMFT supports any arbitrary topology, in the simplest case, the processes can be assumed to be connected in a virtual ring topology. Each process $p_i$ uses the memory of adjacent processor $p_{i+1}$ for checkpointing its local FP-Tree and transactions. Therefore, each process $p_{i+1}$ should prepare its checkpoint buffers ( {\em FPT checkpoints} and {\em transaction checkpoints} vectors) to handle the data checkpointed by process $p_{i}$, when needed during recovery.

To perform a single checkpoint, each pair of processes $(p_{i}, p_{target})$ need to perform three operations. First, $p_{target}$ increases the size of the $metadata_{target}$ and $FPT.chk_{target}$ data structure, such that the new checkpoint from $p_{i}$ can be handled.  The operation of determining the size of the checkpointed $p_{i}$ local FP-Tree requires synchronization between $p_i$ and $p_{target}$. Specifically, $p_{i}$ sends a checkpointing request to $p_{target}$ including the volume of data to be checkpointed. $p_{target}$ uses \texttt{MPI\_Win\_create\_dynamic} mechanism to increase the size of the checkpointed space. The new virtual address is communicated to $p_{i}$, which is used by $p_{i}$ for checkpointing the actual data using \texttt{MPI\_Put} operation.

A process $p_{i}$ may also checkpoint its remaining local transactions on $p_i$ memory to avoid reading it from disk in the case of failure.
If the fault occurs before checkpointing the transactions, remaining transactions are recovered from the disk. However, if $p_{i}$ fails after dataset transactions have been checkpointed, they can be redistributed directly by $p_{target}$ to other available processes. Transactions checkpointing can be performed similar to FP-Tree checkpointing on $Trans.chk_{target}$ vector of the target process.
%
\begin{algorithm} [htp]

\setstretch{1.1}
  \caption{\bf : SMFT FP-Growth Algorithm}
\label{alg:SMFT}
\begin{algorithmic}[1]
\Require {\bf Procedure: initialization($chk\_schema$ = SMFT)}
\newline
\State Create $FPT.chk_i$, $Trans.chk_i$ and $metadata_i$ vectors on $P_i$ (initially-empty).
\State Expose $FPT.chk_i$, $Trans.chk_i$ and $metadata_i$ addresses for read/update //using MPI-RMA.
\NoNumber\hrulefill

\NoNumber {\bf Procedure: performLFPChk ($L.Tree$)}
\newline
\setcounter{ALG@line}{0}
\State Synchronize with $P_{src}$ to resize the $FPT.chk_i$ vector.
\State Add ($L.FPTree$, $FPT.chk_{target}$) \enspace ({\bf MPI\_Put})
\State Update $metadata_{target}$ vector \enspace ({\bf MPI\_Put})

\NoNumber\hrulefill
\NoNumber {\bf Procedure: performTransChk ($L.Trans$)}
\newline
\setcounter{ALG@line}{0}
\State Synchronize with $P_{src}$ to resize the $Trans.chk_{target}$ vector.
\State Add ($Remaining Trans.$, $Trans.chk_{target}$) \enspace ({\bf MPI\_Put})
\State Update $metadata_{target}$ vector \enspace ({\bf MPI\_Put})

\NoNumber\hrulefill
\NoNumber {\bf Procedure: performRecovery ($p_f$, $G.Freq.List$, $P_{rec}$)}
\newline
\setcounter{ALG@line}{0}
\State \emph{$P_{rec}$ process:} merge ($L.Tree$, $P_f.chkFPTree_{rec}$, $G.Freq.List$)
\If {$Trans. chk$ is NULL}
\State   diskTransRec($metadata_{rec}$)
\Else
\State   memTransRec($Trans.chk_{rec}$, $metadata_{rec}$)
\EndIf
\end{algorithmic}
\end{algorithm}

Algorithm~\ref{alg:SMFT} shows the checkpointing and recovery algorithms for SMFT. In $initialization$ procedure, each process create three vectors $FPT.chk_i$, $Trans.chk_i$  and $metadata_i$ vectors  to handle proceeding process checkpoints (Line 1). These vectors are allocated and exposed using MPI-RMA technology for facilitating remote read/update (Line 2).  Both $PerformLFPChk$ procedure and $PerformTransChk$ procedures,  illustrate checkpoint operation in SMFT for both local FP-Tree and transactions, respectively. Process $p_i$ synchronizes with its source process $p_{src}$ by receiving its checkpoint size and resizing its checkpoint buffer to handle $p_{src}$ data. Process $p_i$ finalizes the synchronization operation by sending the new checkpoint vector address to the source process (Line 1). Next,  process $p_i$ uses \texttt{MPI\_Put} function to checkpoint its data  and updates the {\em metadata} vector on target process memory (Lines 2-3).

The $performRecovery$ procedure shows the recovery algorithm in SMFT. The predetermined recovery process $p_r$ is used to recover failed process $P_f$ by merging checkpointed local FP-Tree of $P_f$ it has on its memory to local FP-Tree (Line 1). Further, failed process transactions can be recovered with the aid of {\em metadata} vector directly from {\em recovery process} memory if available or from the disk if not (Lines 2-6). Disk-based recovery should be performed in parallel to speed-up the total recovery time.

\subsection{Asynchronous Memory-based Fault Tolerant (AMFT) FP-Growth}

In the SMFT approach, we observed the advantages of using in-memory
checkpointing of FP-Tree and database transactions. However, there are a
few limitations of SMFT. Specifically, a pair of processes need to
synchronize for memory allocation and address exchange --- which reduces
the overall effectiveness of the MPI One-sided model.

We address the limitations of SMFT by proposing a {\em truly} one-sided
mechanism for checkpointing, i.e., Asynchronous Memory-based Fault Tolerant (AMFT). Under AMFT, we use the memory of {\em
already processed} transactions for checkpointing instead of allocating new space.
Similar to SMFT, under the AMFT approach, it is possible to checkpoint
the FP-Trees and a portion of the database transactions. We describe the
checkpointing, recovery and implementation details of the AMFT approach
as follows.

\noindent
{\bf Checkpointing Algorithm:}
Consider a subset of two processes $\in P$ --- $p_i$ and $p_{target}$. The checkpoint from $p_{i}$ is stored on $p_{target}$. To enable truly one-sided mechanism for checkpointing, $p_{i}$ must ensure that its checkpoint size is less than the size of the already processed transactions in $p_{target}$. In AMFT, we achieve this objective by using atomic operations on variables  allocated using MPI-RMA and exposing it to read/update by other processes. The original parallel FP-Growth algorithm is slightly modified to atomically update the size of available checkpointing space --- this step does not require communication with any other process. When $p_{i}$ decides to checkpoint its FP-Tree, it atomically reads
the value of available checkpointing space on $p_{target}$. By carefully designing the checkpointing interval, it is highly likely that the size of the available checkpointing space on $p_{target}$ is greater than the size required by $p_{i}$. In the pathological case, $p_{i}$ periodically reads the available checkpointing space, till the condition is satisfied --- in practice, this situation is not observed. In the common case, $p_{i}$ simply initiates the checkpoint using \texttt{MPI\_Put}. Besides local FP-Tree, remaining (unprocessed) transactions of process $p_i$ can also be checkpointed to $p_{target}$ memory if there is enough space. Checkpointing transactions is one-time operation that improves the recovery process by reading failed process's transactions directly from checkpoint memory space instead of disk.

\begin{figure}[htp]
\centering
  \subcaptionbox{\centering \small Local FP-Tree Checkpointing\label{fig:b0}}{\includegraphics[width=1.2in,height=1.75in]{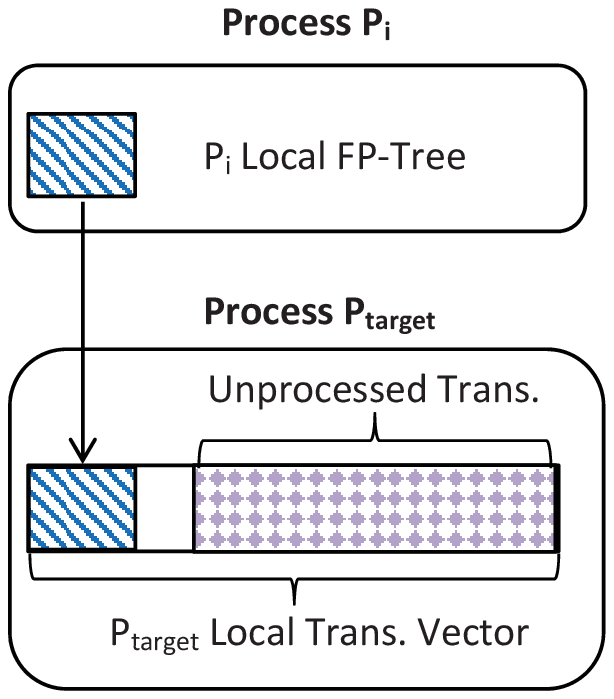}}\hspace{1.5em}%
  \subcaptionbox{\centering \small Unprocessed Transactions and Local FP-Tree Checkpointing\label{fig:b1}}{\includegraphics[width=1.4in,height=1.9in]{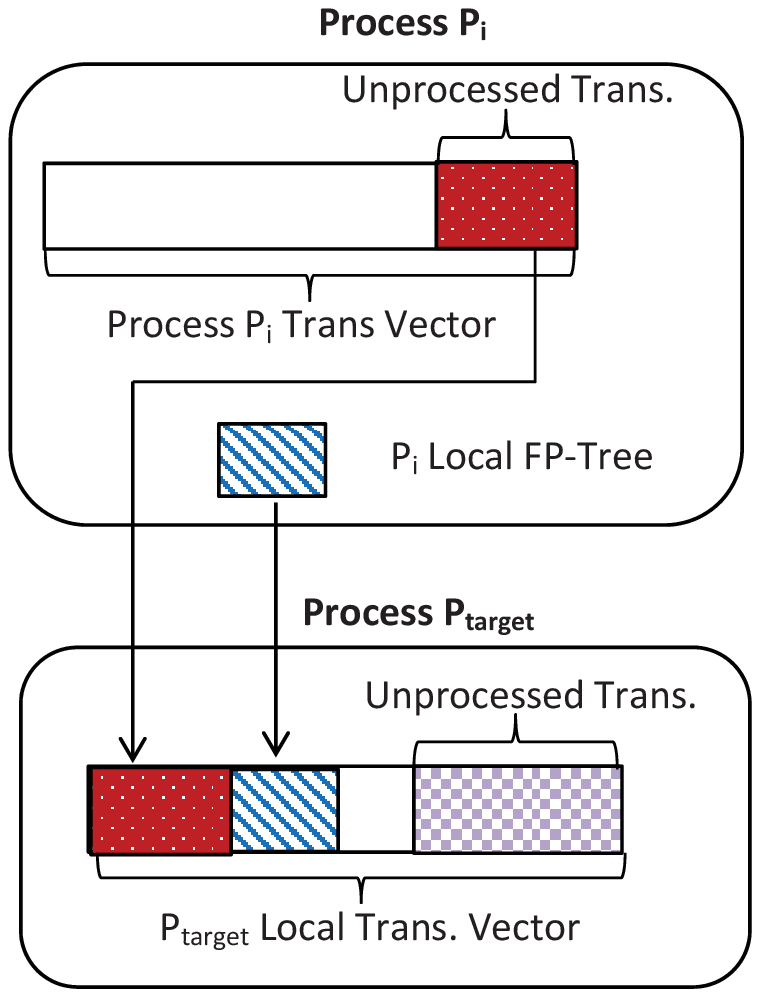}}\hspace{0em}%
   \caption{AMFT Checkpointing Operation Overview}
  \label{fig:AMFT-chk}
\end{figure}

Figure~\ref{fig:AMFT-chk} illustrates AMFT checkpointing operation by showing two different cases. In Figure~\ref{fig:b0} only local FP-Tree of process $p_i$ is checkpointed on $p_{target}$ available transactions space. However, in Figure~\ref{fig:b1} both remaining transactions and local FP-Tree of process $p_i$ are checkpointed to $p_{target}$ memory (i.e., memory space availability is required).

The effectiveness of AMFT checkpointing algorithm is in its simplicity. Unlike SMFT, there is no synchronization required between any pair of processes, and memory allocation is not required as well. By using MPI-RMA on high performance interconnects such as InfiniBand, we expect AMFT to be a near-optimal checkpointing algorithm for designing large scale FP-Growth algorithm. As expected, since each process simply initiates the communication for the checkpoint, the expected time
complexity of the checkpointing is $O(\frac{|T|}{log |P|.C})$, using the LogGP model~\cite{alexandrov:95}.

\noindent
{\bf Recovery Algorithm:}
The recovery algorithm for AMFT is similar to SMFT. Assuming $p_{target}$ is the recovery process $p_{rec}$. When a fault occurs (on $p_i$), {\em recovery process} $p_{rec}$  merges the checkpointed FP-Tree of $p_i$ with its FP-Tree and re-distributes the dead process $p_i$ transactions among a subset of available processes (such as $\log{|P|}$), if an in-memory checkpoint is available locally.  Otherwise, all available processes recovered unprocessed transactions of the failed process $p_i$ from the disk in parallel.

The worst case time complexity of AMFT approach is similar to SMFT. In the worst case, the entire transactions are read from disk in parallel as mentioned in SMFT approach with ($\frac{|T|}{|P|\cdot |P-1| \cdot l}$) time complexity, and recomputed by $\log{|P|}$ processes in ($\frac{|T|}{|P|\cdot \log{|P|}}$). However, in many cases --- especially when the fault occurs during later stages of FP-Tree build phase --- disk will be completely avoided, resulting in much faster recovery in comparison to the worst case scenario.


\noindent
{\bf Implementation Details:}
\begin{algorithm} [htp]
\setstretch{1.1}
  \caption{\bf : AMFT FP-Growth Algorithm}
\label{alg:AMFT}
\begin{algorithmic}[1]
\Require {\bf Procedure: initialization($chk\_schema$ = AMFT)}
\newline
\State Assume $Trans_i$ vector is the memory space contains $L.Trans$ on $P_i$
\State Create $metadata_i$ vector on $P_i$ to describe $P_{src}$ checkpoint (initially-empty).
\State Expose $Trans_i$ and $metadata_i$ vectors addresses for read/update  \enspace //using MPI-RMA

\NoNumber\hrulefill
\NoNumber {\bf Procedure: performChk ($L.FPTree$, $L.Trans$, $P_{target}$ )}
\newline
\setcounter{ALG@line}{0}
\If {$Trans_{target}$  has enough space for $L.FPTree$ of $P_i$}
\State  add($L.FPTree$, $Trans_{target}$) ( \enspace ({\bf MPI\_Put})
\EndIf
\If {$Trans_{target}$  has enough space for remaining $L.Trans$ of $P_i$ (Only one time)}
\State  add($L.Trans$, $Trans_{target}$)  \enspace ({\bf MPI\_Put})
\EndIf
\State Update $metadata_{target}$ vector  \enspace ({\bf MPI\_Put})

\NoNumber\hrulefill
\NoNumber {\bf Procedure: performRecovery ($P_f$, $G.Freq.List$, $P_{rec}$ )}
\newline
\setcounter{ALG@line}{0}

\State  \emph{$P_{rec}$ process:}  merge ($L.Tree$, $P_f.chkFPTree$, $G.Freq.List$)
\If {$Trans. checkpoint$ is NULL}
\State   diskTransRecv($metadata$)
\Else
\State   memTransRecv($Trans$, $metadata$)
\EndIf
 \end{algorithmic}
\end{algorithm}

Algorithm~\ref{alg:AMFT} illustrates the checkpointing and recovery procedures for AMFT algorithm. During the initialization procedure, each process has its own $Trans_i$ vector that contains local set of transactions $L.Trans$ (Line 1). In line 2, each process $p_i$ creates a single vector, i.e., $metadata_i$,  that represents a set of parameters to describe the status of $L.Trans$ vector and checkpointed data of source process $p_{src}$ stored on $p_i$ memory. In line 3, MPI-RMA technology is used to shared both vectors, i.e., $Trans_i$ and $metadata_i$, to other processes.

Both $L. FPTree$ and remaining transactions $L.Trans$ can be checkpointed using $performChk$ procedure. Each process should read $metadata_{target}$ on target process $p_{target}$ to check for space availability before checkpointing (Lines 1-6).  Remaining transaction $L.Trans$ checkpointing is only performed one time once a space is available.

The $performRecovery$ procedure shows the recovery algorithm in AMFT approach. Like the SMFT recovery algorithm, the recovery process $P_r$ process is used to recover $p_f$ by merging latest checkpointed  FP-Tree $p_f$ it has with its local FP-Tree. $p_f$ unprocessed transactions can be recover from {\em recovery process} memory if it was checkpointed before failure or directly from disk (Lines 2-6).

  \begin{figure*}[htp]
\centering
  \subcaptionbox{100M Trans. $\theta$=$0.05$\label{fig:x0}}{\includegraphics[width=1.7in]{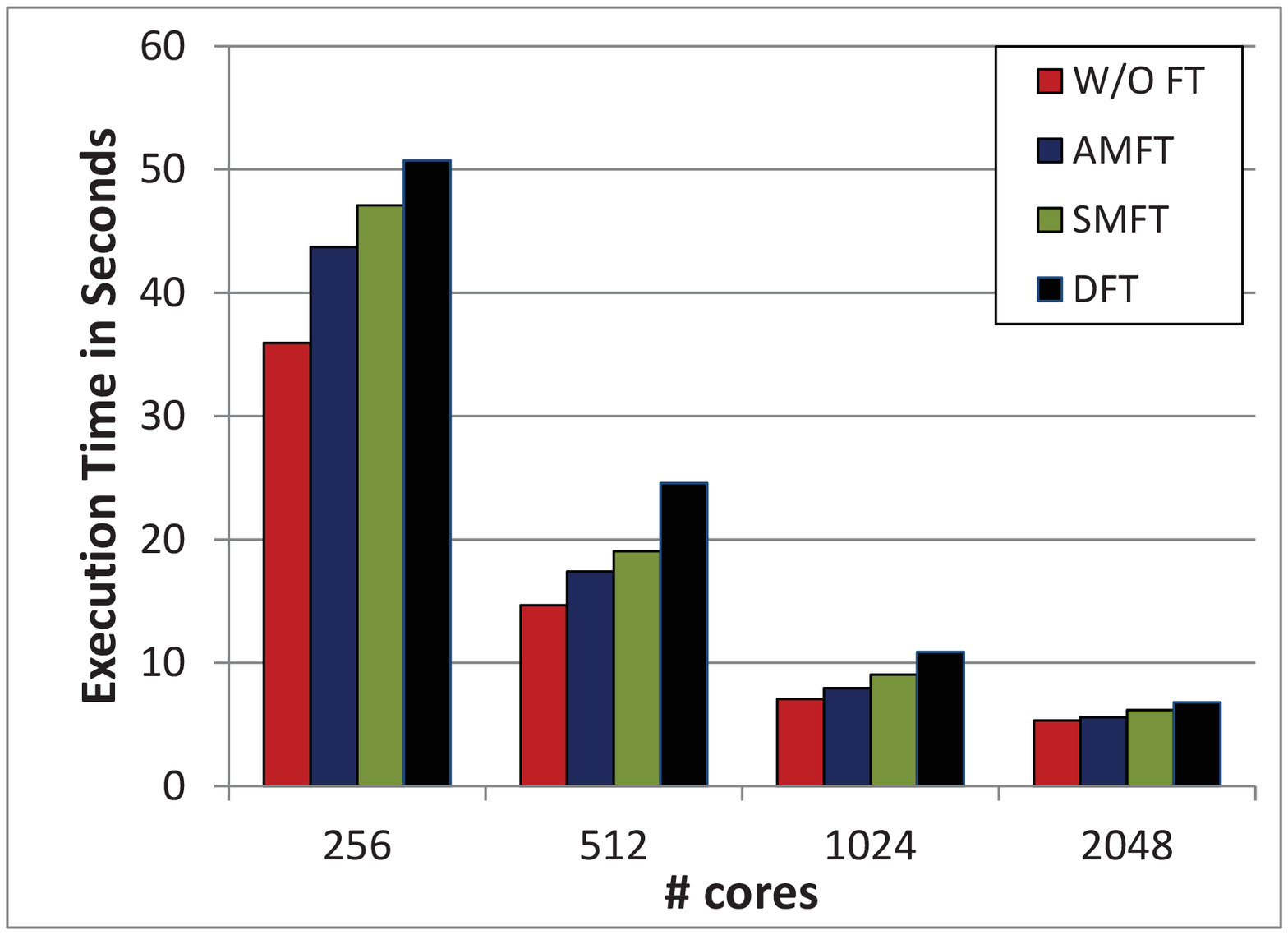}}\hspace{0.07em}%
  \subcaptionbox{100M Trans. $\theta$=$0.03$\label{fig:x1}}{\includegraphics[width=1.7in]{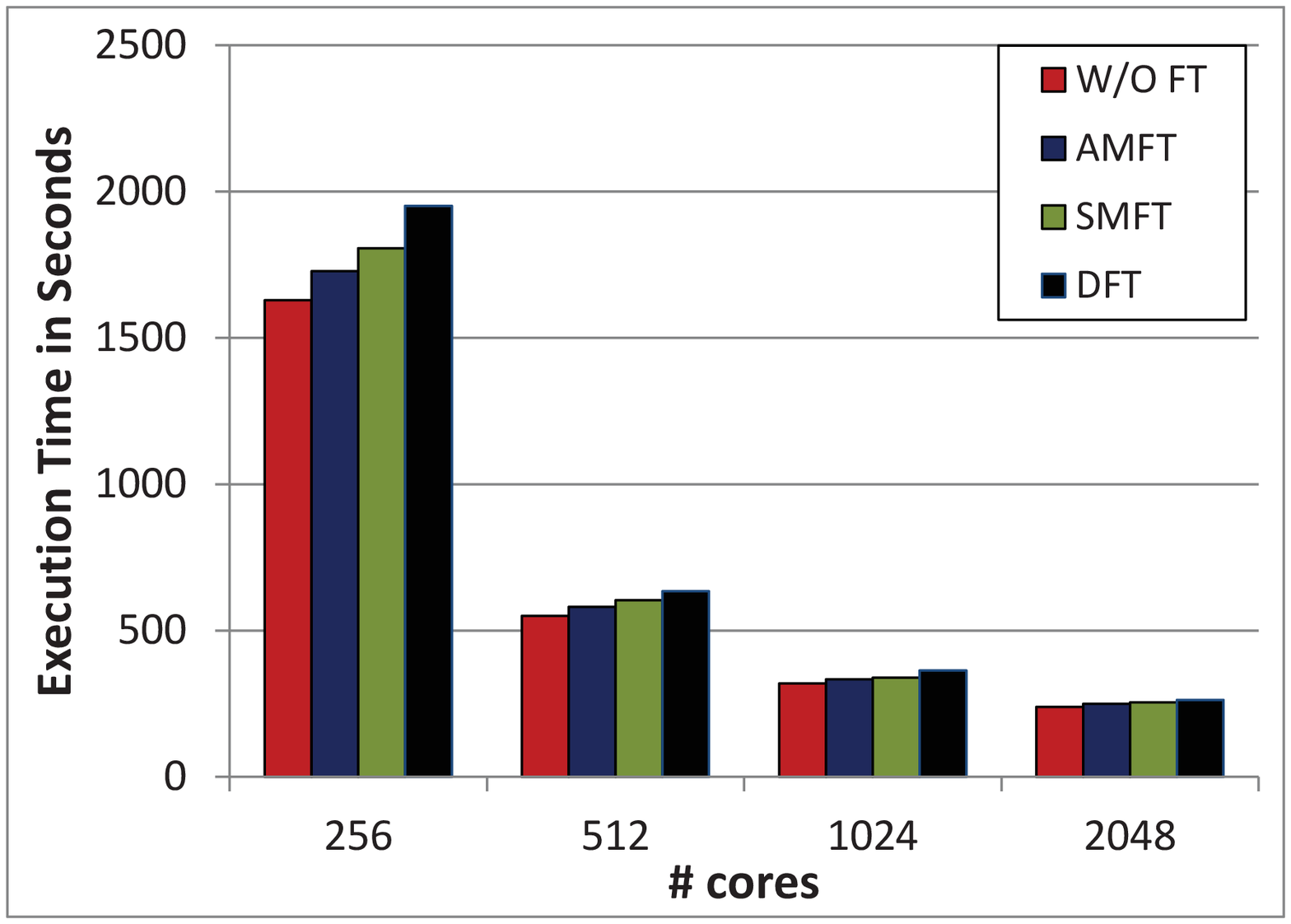}}\hspace{0.07em}%
   \subcaptionbox{200M Trans. $\theta$=$0.05$\label{fig:x2}}{\includegraphics[width=1.7in]{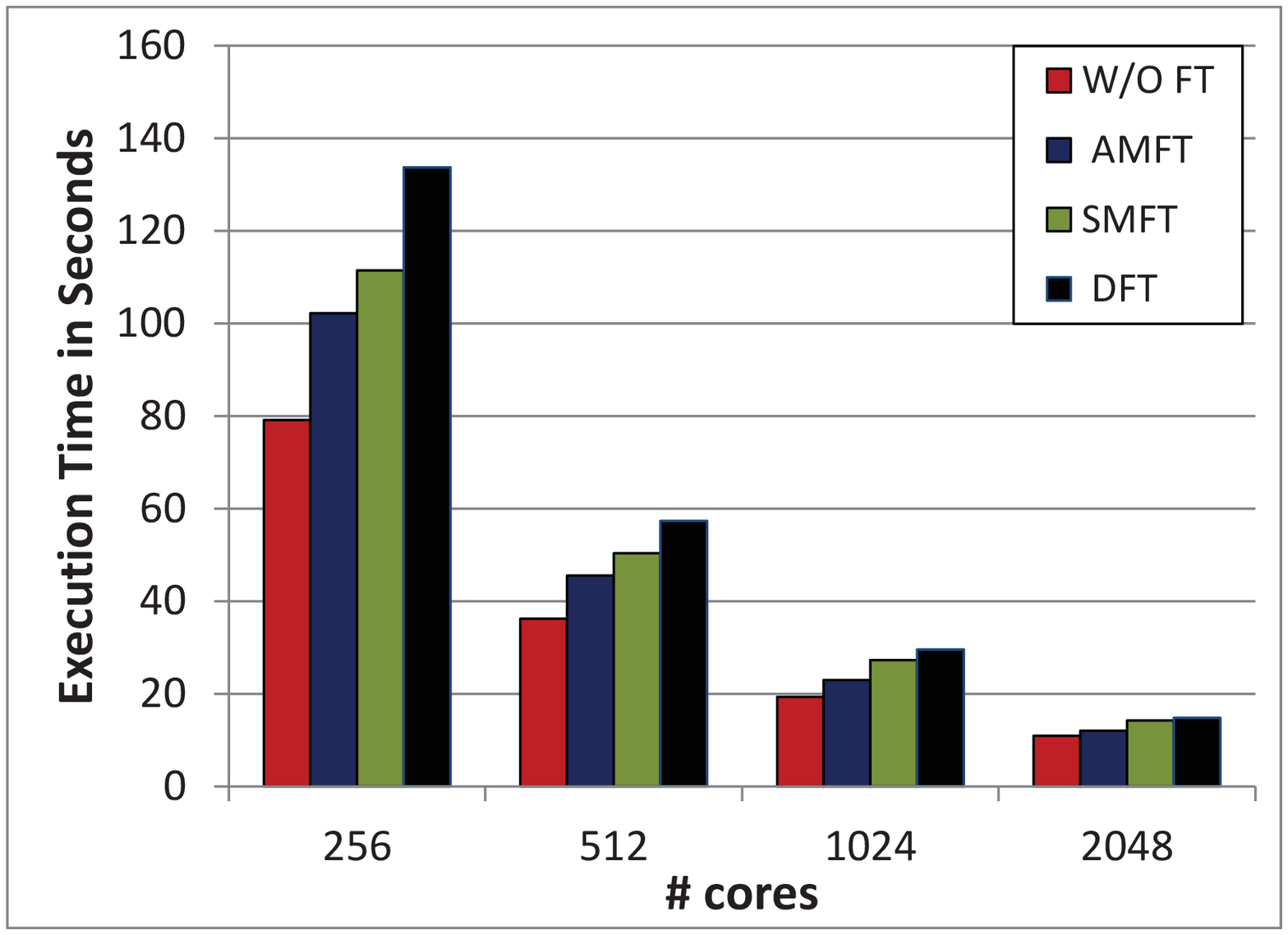}}\hspace{0.07em}%
  \subcaptionbox{200M Trans. $\theta$=$0.03$ \label{fig:x3}}{\includegraphics[width=1.7in]{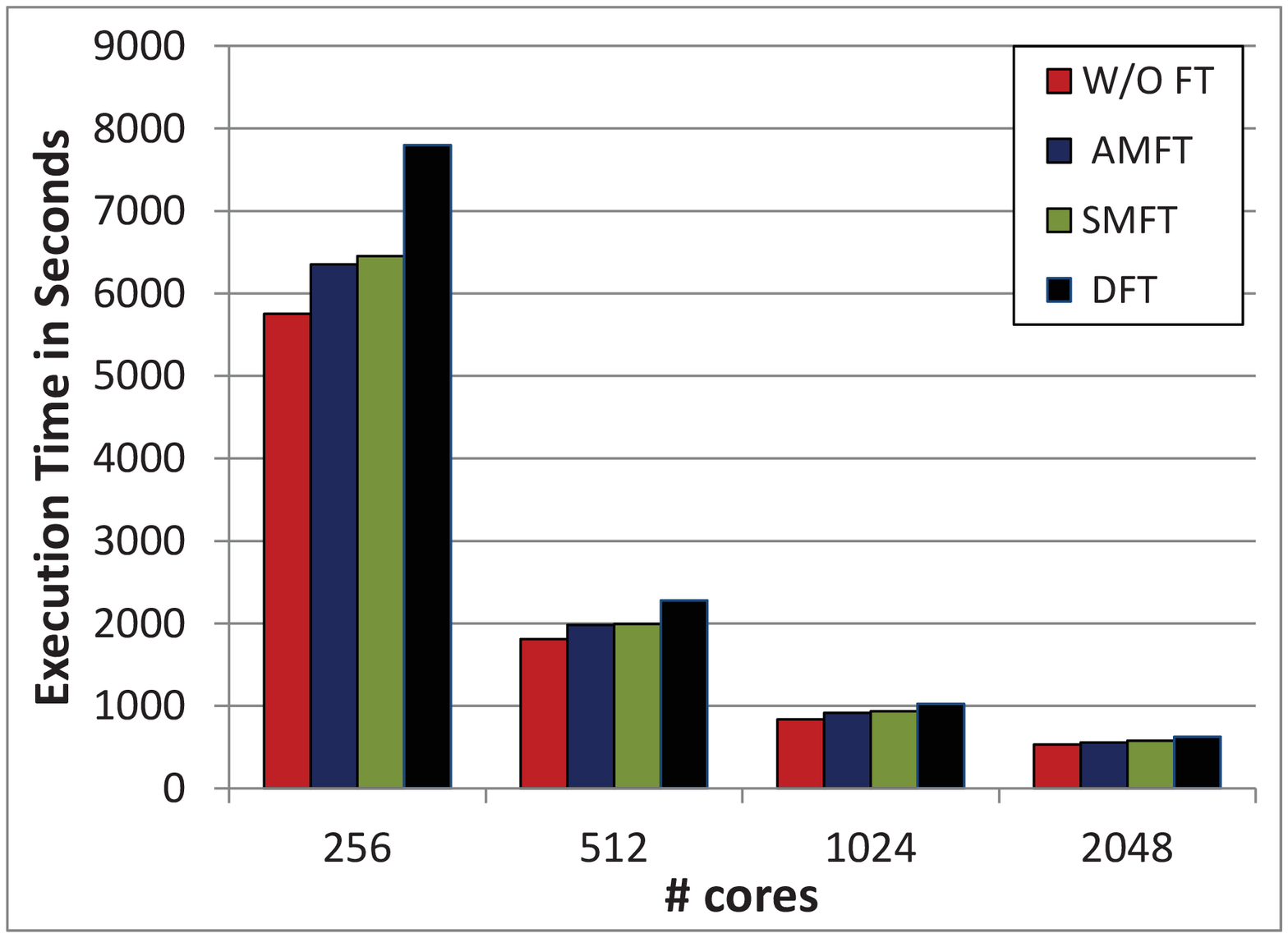}}\hspace{0.07em}%

   \caption{Proposed FT mechanisms checkpointing overhead  with different number of transactions, support threshold, and cores}
  \label{fig:uniformdataset1}
\end{figure*}
\section{Performance Evaluation}
\label{sec:perf}
In this section, we present a detailed performance evaluation of the proposed fault tolerant FP-Growth algorithms, i.e.,  DFT, SMFT, and AMFT  that were presented in section~\ref{sec:design}. For each fault tolerant algorithm, we present a detailed performance analysis of the checkpointing and recovery overhead. We use up to 200 million transactions and a large scale evaluation using up to 2048 cores. At the end of this section, a comparison against a fault-tolerant version  executed on Spark is  presented.

\subsection{Setup}
\subsubsection{Experimental Testbed}
We use Stampede supercomputer at the Texas Advanced Computing Center (TACC) for performance evaluation. The Stampede supercomputer is Dell PowerEdge C8220 cluster with 6,400 Dell PowerEdge server nodes, each with 32GB memory, (2) Intel Xeon E5 (8-core Sandy Bridge) processors. We use  MVAPICH2-2.1, a high performance MPI library available on Remote Direct Memory Access (RDMA) interconnects such as InfiniBand.  We use aggressive compiler optimizations with Intel compiler v15.0.1 for performance evaluation.

\subsubsection{Datasets}
To evaluate  different proposed fault tolerant FP-Growth algorithms, we use IBM Quest dataset generator~\cite{agrawal2009quest} for generating large scale synthetic datasets. IBM Quest dataset generator has been widely used in several studies, and accurately reflects the pattern of transactions in real-world datasets~\cite{yu2011efficient,buehrer2006out,xu2014efficient,lin2014determining}. For experimental evaluation, we use two synthetic datasets with 100 and 200 million transactions.  The number of items per transaction is 15-20. A total of 1000 item-ids are used. 
\subsection{Overhead of Supporting FP-Growth Fault Tolerance}
 \subsubsection{Checkpointing Overhead Evaluation}
While the recovery algorithm is executed only during faults, the cost of checkpointing is incurred even in the absence of faults. Naturally, it is critical to minimize the checkpointing time --- especially, when the fault rates are low.

\tabcolsep=0.11cm
\begin{table}[!h]
\singlespacing
\centering
\caption{DFT, AMFT, and SMFT systems slowdowns related to w/o FT FP-Growth algorithm}    
    \label{tbl:slowdowns}
    \tabcolsep=0.15cm
      \begin{small}
    \begin{scriptsize}
    \begin{tabular}{|c|c|c|c|c|c|c|c|}
    \hline
\centering
    {\bfseries \# Cores} &   {\bfseries  Sup.} & \multicolumn{2}{|c}{\bfseries DFT (\%)}  &\multicolumn{2}{|c}{\bfseries SMFT (\%)} & \multicolumn{2}{|c|}{\bfseries AMFT (\%)}\\
\cline{3-8}

    {\bfseries} & {\bfseries} & 100M & 200M & 100M & 200M & 100M & 200M\\

    \hline

    256&0.03&19.76 & 35.59 & 10.85 &12.23 &6.08&10.4 \\

    {\bfseries} & 0.05 &67.31&69.01&31.02&40.8&21.62&29.09 \\
     \hline

    512 & 0.03 &15.28&25.87&9.76&10.01&5.66&9.32\\

    {\bfseries} & 0.05 & 54.11&58.07&29.77&38.88&18.50&25.5	\\
     \hline

    1024 & 0.03 & 13.77&22.22&6.67&11.52&4.47&9.14	\\

    {\bfseries} & 0.05 & 41.17&52.93&27.87&41.34&12.21&18.8\\
     \hline

    2048 & 0.03 & 10.13&17.56&6.05&8.69&4.21&7.9	\\

    {\bfseries} & 0.05 & 27.39&35.61&15.83&30.15&5.11&9.92	\\
     \hline

\end{tabular}
\end{scriptsize}
\end{small}
\end{table}

Figure~\ref{fig:uniformdataset1} shows the checkpointing overhead of
DFT, SMFT and AMFT algorithms using 100M, 200M transactions and
support threshold ($\theta$) values of 0.03 and 0.05. Table~\ref{tbl:slowdowns}
presents the data in a tabular form, by showing the percentage of
slowdown in comparison to the default parallel algorithm that is
not  fault-tolerant.
In  Figure~\ref{fig:uniformdataset1}(a), if we focus on   strong
scaling evaluation (keeping the overall work constant and increasing the
number of processes), the algorithm scales very well
(scaling from 256 -512 processes, we observe super-linear speed-up due
to better cache utilization). Similar speed-ups are observed for DFT,
SMFT,  and AMFT algorithms, respectively. Since the support threshold is
high (0.05), the number of frequent item-ids is relatively small. Hence,
the overall computation time is less than 50s. Naturally, the slow down
observed by DFT and SMFT is high --- 67\% and 31\%, respectively. AMFT
only experiences a slowdown of 21\%. We expected negligible overhead for
AMFT. However, we experienced slowdown, because for small scales such as
256 processes, the size of individual FP-Tree is larger (in comparison
to larger process counts). Unfortunately, current MPI-RMA
implementations are not always optimized for bulk data transfers. To
validate this argument, we observe the column for AMFT with 100M
transactions. On 2048 cores --- with strong scaling --- the overhead of
checkpointing reduced to 5\%. For lower support threshold, as shown in
Figure~\ref{fig:uniformdataset1}(b), the overall slowdown for AMFT is
4-6\%, while DFT overhead is 10-20\%, for different process counts.

Figure~\ref{fig:uniformdataset1}(c) shows the performance comparison of
DFT, SMFT,  and AMFT algorithms using 200M transactions and 0.05 support
threshold. We observe similar pattern as
Figure~\ref{fig:uniformdataset1}(a). While we expect relatively high
overheads for DFT and SMFT approaches, we observe higher relative
overhead for AMFT approach as well. We argue that for larger
transactions per process, the size of the FP-Tree is larger. Since
MPI-RMA runtimes are less optimized for bulk transfer, the slowdown is
smaller, but non-negligible.

\begin{figure*}[htp]
\centering

  \subcaptionbox{100M Trans. $\theta$=$0.05$\label{fig:x5}}{\includegraphics[width=1.6in]{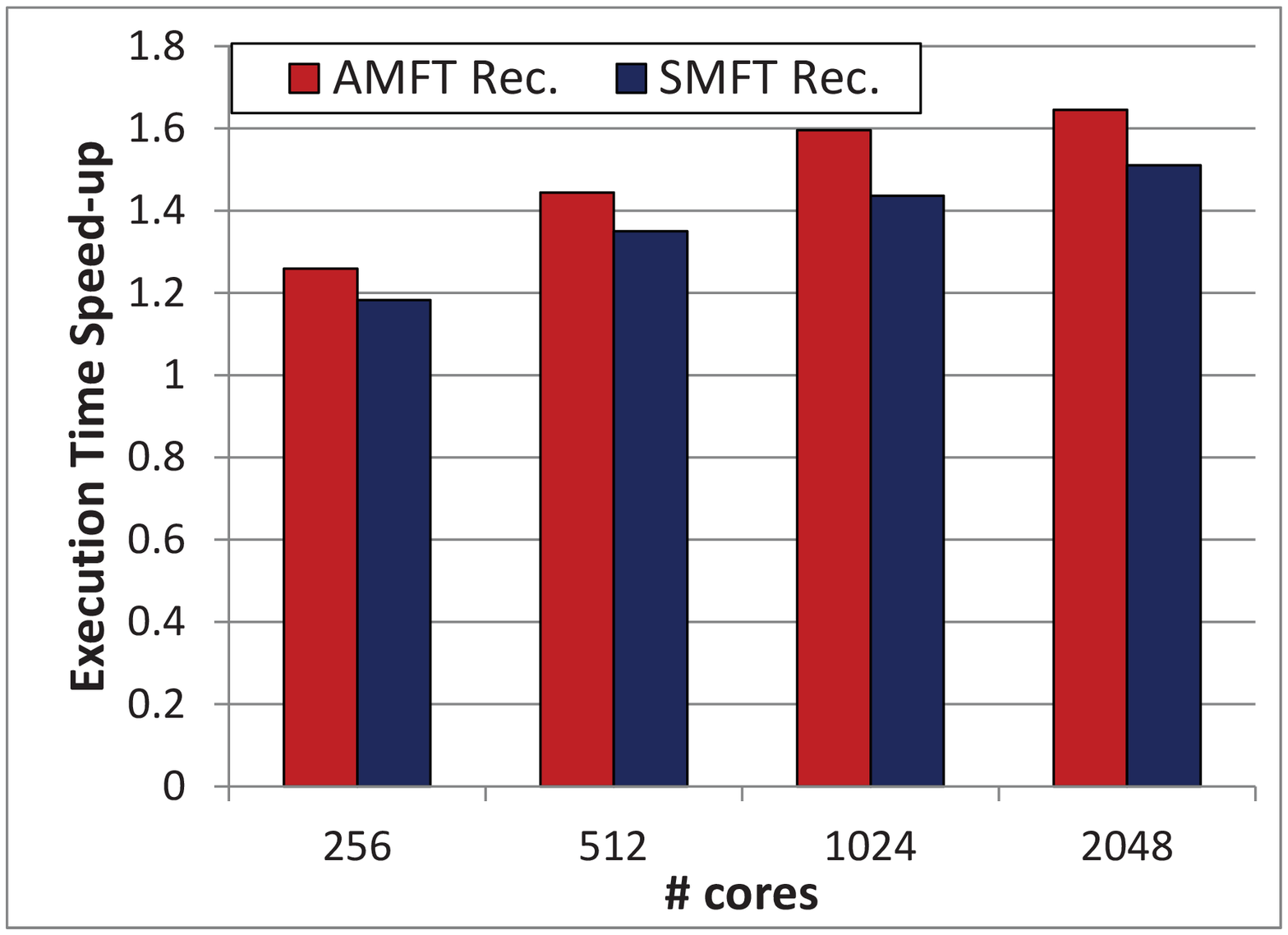}}\hspace{0.07em}%
  \subcaptionbox{100M Trans. $\theta$=$0.03$\label{fig:x6}}{\includegraphics[width=1.6in]{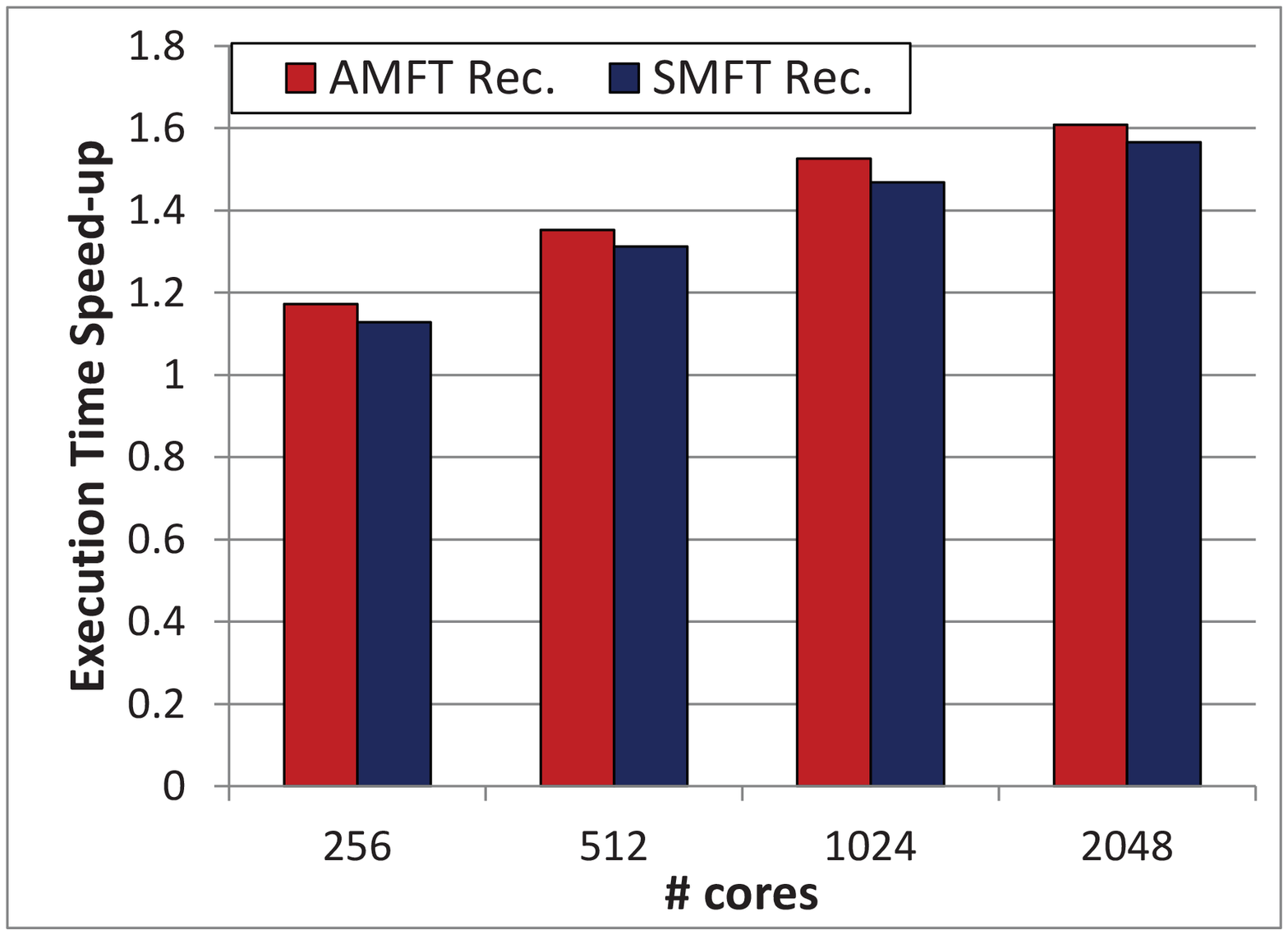}}\hspace{0.07em}%
   \subcaptionbox{200M Trans. $\theta$=$0.05$\label{fig:x7}}{\includegraphics[width=1.6in]{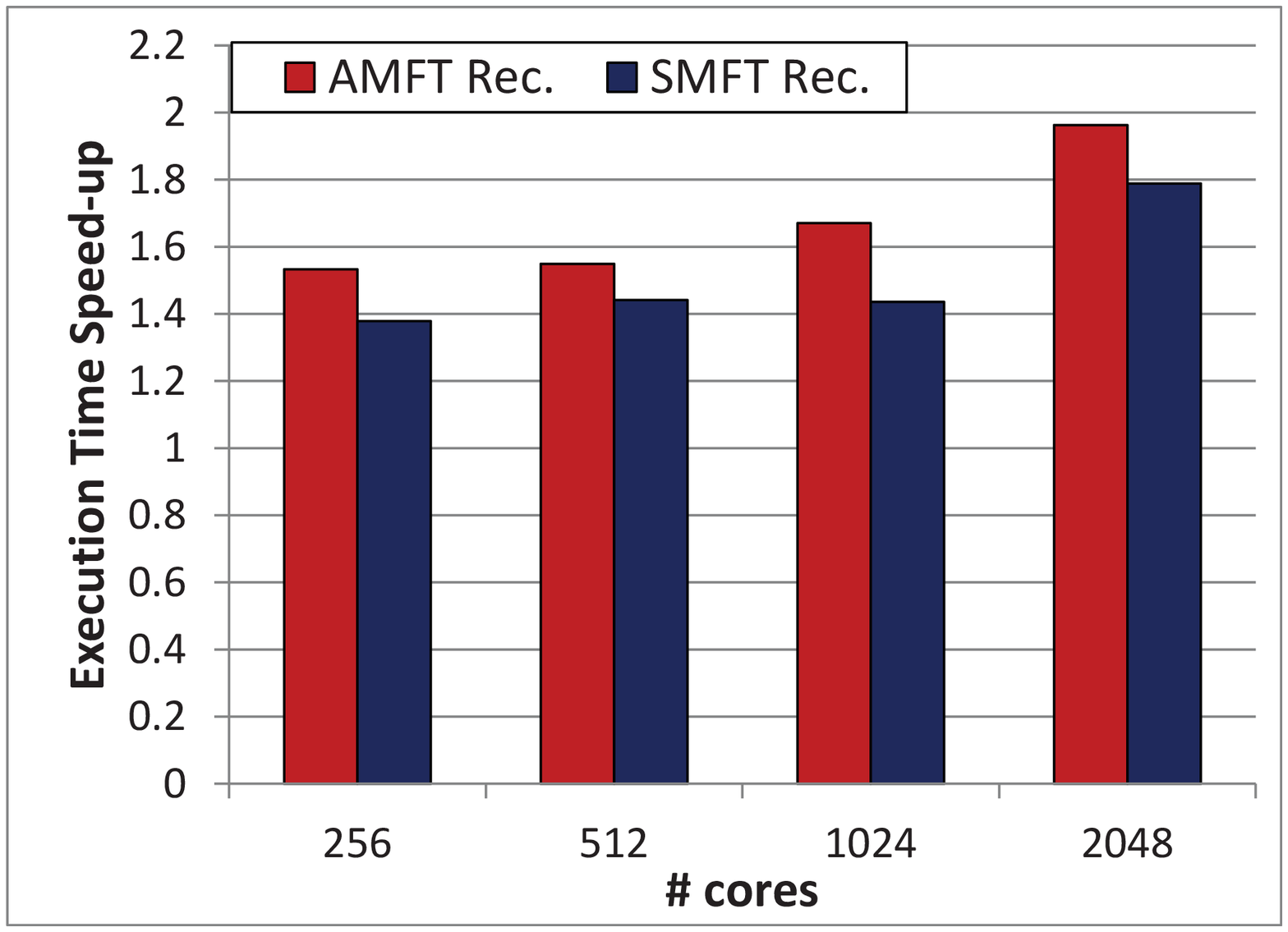}}\hspace{0.07em}%
  \subcaptionbox{200M Trans. $\theta$=$0.03$ \label{fig:x8}}{\includegraphics[width=1.6in]{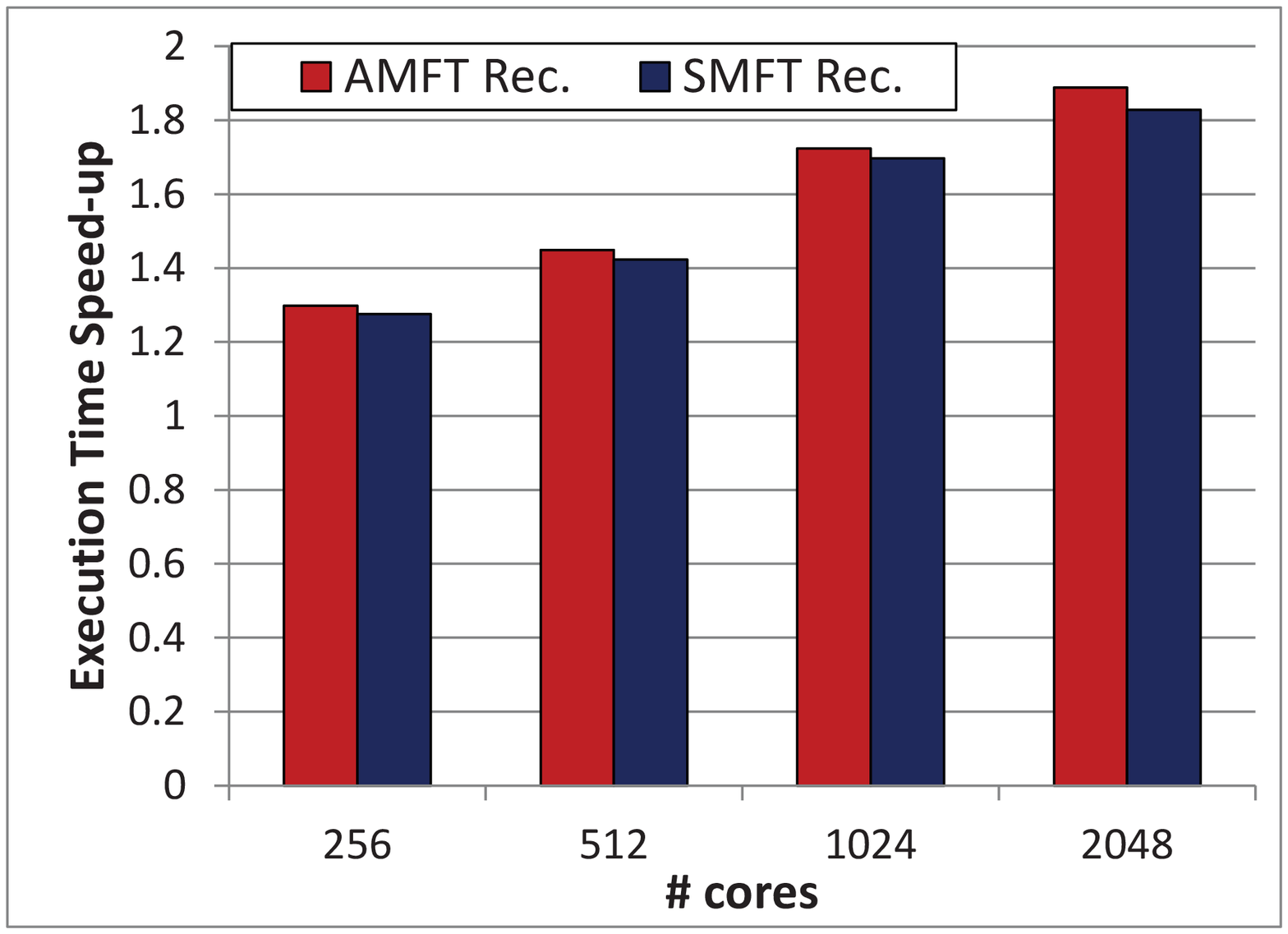}}\hspace{0.07em}%
 \caption{SMFT and AMFT Recovery Speed up Compared to DFT Approach with Different Number of transactions, Support
 Threshold , and Cores}

       \label{fig:recoveryoverhead2}
\end{figure*}

Figure~\ref{fig:uniformdataset1}(d) illustrates the performance of the
proposed approaches with 200M transactions and 0.03 support threshold.
The DFT approach observes a slowdown of 17-35\% in comparison to the
basic parallel algorithm, while AMFT only observes up to 10\% overhead.

Clearly AMFT outperforms other approaches, especially the disk-based approach easily without incurring any additional space complexity. We
also observe that with strong scaling, which is usually a problem for
distributed memory algorithms, the relative overhead of AMFT decreases.
We argue that it is due to the unoptimized MPI-RMA protocols for bulk
data transfer. With further optimizations, as expected in near future,
these overheads are expected to reduce further. With $O(1)$ space
complexity and still acceptable checkpointing overhead such as 10\% for
AMFT, we expect the proposed algorithm to be used as the basis for
future research and practical deployments.


\subsubsection{Recovery Overhead Evaluation}
The effectiveness of any fault tolerance mechanism is related to failure
recovery overhead besides the checkpointing overhead. In this
subsection, we evaluate the recovery overhead in the case of failure by
injecting faults into FP-Growth parallel execution.  To simulate faults,
we select a process to fail and the point of failure.  When
reaching failure point, that process is eliminated from the execution.
We assume failure point after processing 80\% of dataset transactions to
fairly comparing recovering algorithm for DFT, SMFT , and AMFT
approaches.

In the case of failure, DFT recovery algorithm needs to recover FP-Tree of failed
process from the disk comparing to both SMFT and AMFT approaches
where FP-Tree is recovered from memory. In the first set of experiments,
we calculate the speed-up using both SMFT and AMFT approaches
compared to DFT approach to recovery one failure process as shown in
Figure~\ref{fig:recoveryoverhead2}.  In Figures~\ref{fig:x5} and
~\ref{fig:x7}, with 0.05 support threshold, the average speed-up
by SMFT algorithm is  1.36x while average gained speed-up  by AMFT algorithm is 1.41x using
100M dataset in the recovery process. In the case of 200M synthetic dataset,
both SMFT and AMFT recovery algorithms speed-up the total execution time with recovery by 1.55x and 1.59x, respectively, compared to DFT algorithm. In Figure~\ref{fig:x6} and~\ref{fig:x8}, with 0.03 support threshold, the recovered FP-Tree becomes
larger which negatively impacts the performance of DFT approach compared
to the other two approaches (i.e., SMFT and AMFT). Thus, with 100M dataset,
compared to DFT approach, SMFT  speeds-up the recovery process by 1.39x
while AMFT speeds-up the recovery process with 1.46x. Using 200M
dataset, SMFT speeds-up the algorithm execution with recovery by 1.51x while AMFT
speeds-up the algorithm with 1.68x.

\tabcolsep=0.15cm
\begin{table}[!h]
\singlespacing
\centering
\caption{DFT, SMFT and AMFT Total Execution Time Including The  Recovery Time}    
    \label{tbl:speed-up}
    \tabcolsep=0.15cm
      \begin{small}
    \begin{scriptsize}
    \begin{tabular}{|c|c|c|c|c|c|c|c|}
    \hline

    {\bfseries \# Cores} &   {\bfseries  Sup.} & \multicolumn{2}{|c}{\bfseries DFT Time (Sec)} & \multicolumn{2}{|c}{\bfseries SMFT Time (Sec)}&
     \multicolumn{2}{|c|}{\bfseries AMFT Time (Sec)}\\
                          \cline{3-8}
{\bfseries} & {\bfseries} & {\bfseries 100M} & {\bfseries 200M}&{\bfseries 100M} & {\bfseries 200M}&{\bfseries 100M} & {\bfseries 200M}\\

 \hline

    256&0.03&2312.65 & 8860.26 & 2049.68& 6945.23&1972.01	& 6822.59 \\
    \cline{2-8}
    {\bfseries} &0.05 & 67.12 & 182.685 &56.57 &132.52 &  54.23 & 119.16\\
     \hline

      512 & 0.03 & 948.125&	3227.25&	722.19&	2268.12&	701.12&2226.65\\
    \cline{2-8}
    {\bfseries} & 0.05 & 34.59&	92.36	&26.95&	64.12	&24.83&59.63\\

     \hline

         1024 & 0.03 & 609.52&1762.34&	415.12	&1038.23&	399.52&1022.52\\
    \cline{2-8}
    {\bfseries} & 0.05 & 15.88&	45.48	&11.06	&31.68&	9.95&27.23\\

     \hline

       2048 & 0.03 & 438.85	&1151.12&	280.23	&629.62&272.85&609.62\\
    \cline{2-8}
    {\bfseries} &0.05& 10.55 & 27.04	&6.97&15.12	&6.40	&13.78\\

     \hline

\end{tabular}
\end{scriptsize}
\end{small}
\end{table}

Table ~\ref{tbl:speed-up} summarizes the total execution time including the recovery time of DFT, SMFT, and AMFT algorithms to handle one failure using 256,
512, 1024, and 2048 cores with 0.03 and 0.05 support threshold,
respectively.  Several observations can be drawn from Figure~\ref{fig:recoveryoverhead2} and
Table~\ref{tbl:speed-up}. Both SMFT and AMFT algorithms  speed-up the
FP-Growth algorithm recovery process compared to DFT algorithm. With smaller support threshold ($\theta$=0.05), the size of checkpointed
local FP-Trees and dead process recovered FP-Tree is small. Thus, in
SMFT the synchronization overhead can be clearly shown compared to AMFT
algorithm. In this case, AMFT outperforms SMFT algorithm as shown. However,
in the case of ($\theta$ =0.03), the size of FP-Tree is larger and the
synchronization overheads are small compared to checkpointing and recovery
time. Thus, the speed-up difference between SMFT and AMFT decreases.
Another observation that could be obvious is that the average speed-up for
both SMFT and AMFT algorithms increases with larger dataset (i.e.,
200M). The main reason of this that FP-Trees become larger and DFT
algorithm needs more time to checkpoint  or recover it from disk.
Finally,  with ($\theta$ =0.3), we observe a super-linear speed-up from 256 to 512 cores due to better cash utilization.

\subsection{Comparison Against Spark}
We compare our proposed AMFT FP-Growth algorithm with Spark FP-Growth algorithm to show the effectiveness of our proposed system.
Although, it is common for MPI-based  implementations   to outperform MapReduce-based implementations~\cite{hoefler2009towards},
we are particularly interested in absolute and relative overheads for handling failures.  Spark has a built-in Machine
Learning library (MLlib)  that  includes  an FP-Growth algorithm, which   we use in our comparison. A set of experiments has been conducted with different number of nodes and using 500K synthetic dataset to show the performance of both MPI-based and spark-based FP-Growth algorithms.

\begin{figure}[!htb]
\centering
 \subcaptionbox{500K Trans. $\theta$=$0.03$
  \label{fig:z5}}
 {\includegraphics[width=42mm]{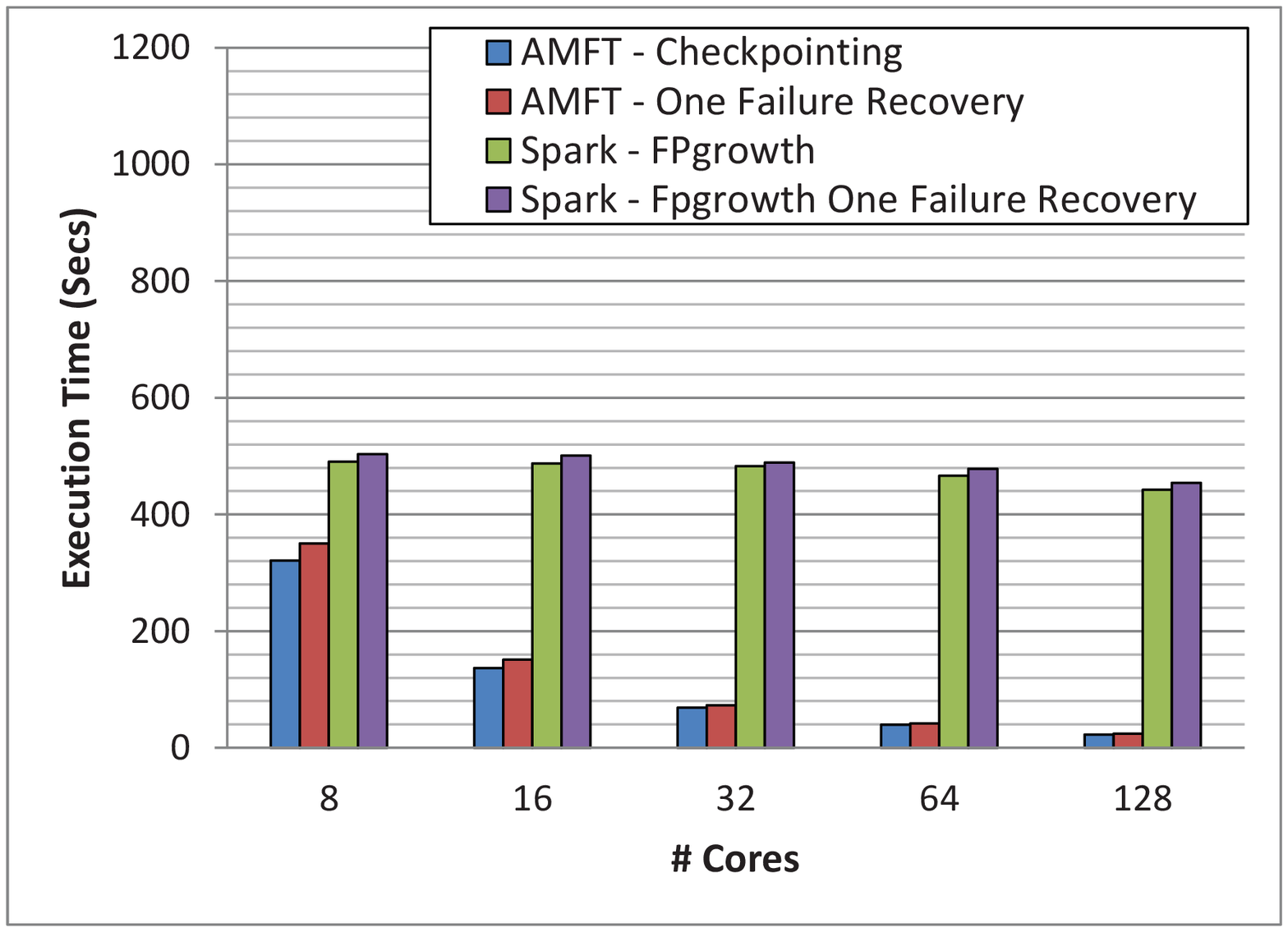}}
   \subcaptionbox{500K Trans. $\theta$=$0.01$
  \label{fig:z6}}
 {\includegraphics[width=42mm]{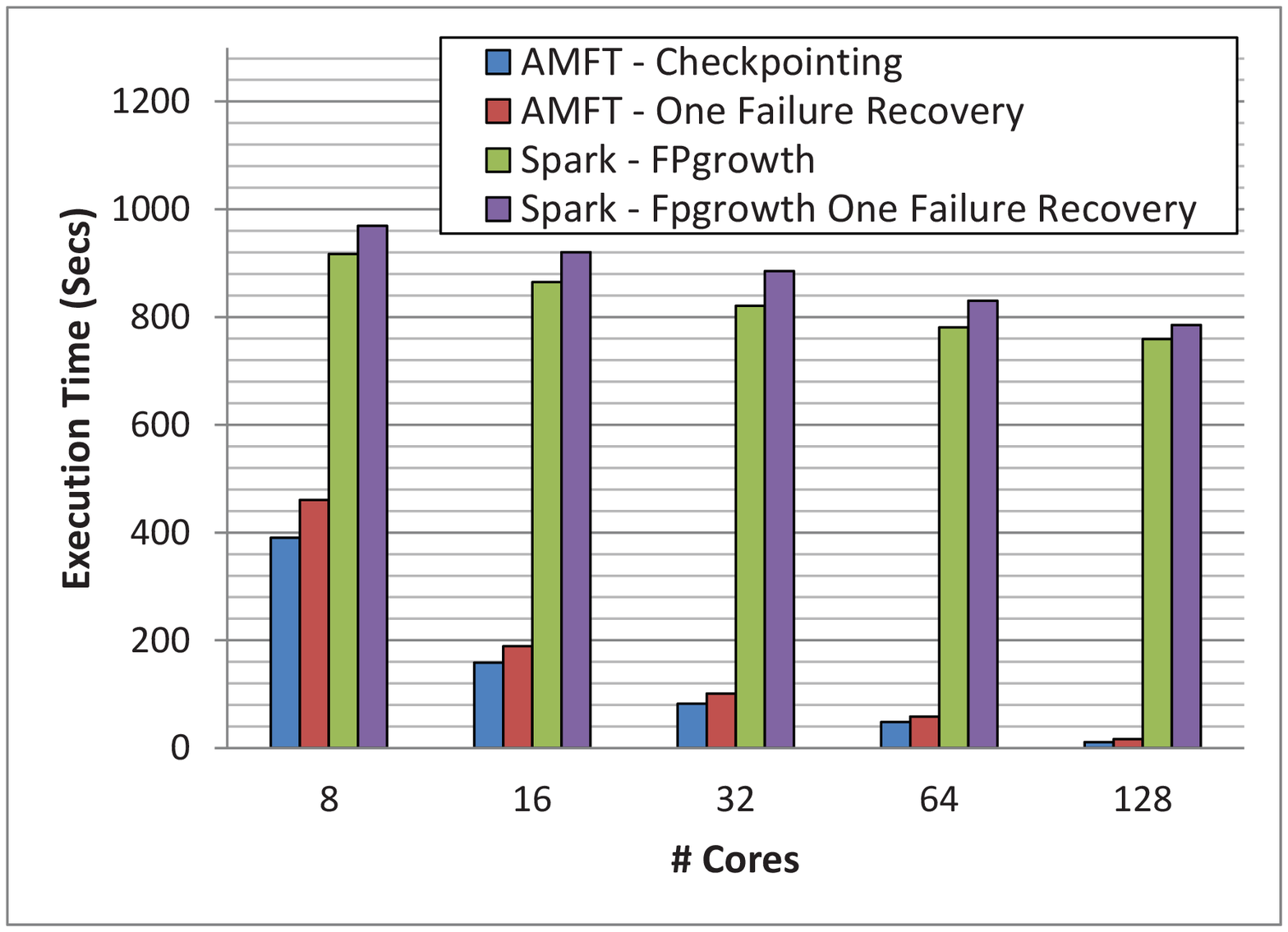}}

\caption{Spark and MPI-based (AMFT))with Different Support Threshold $\theta$ and using 500K Synthetic Dataset}
 \label{fig:spark-fp}
\end{figure}

Figure~\ref{fig:spark-fp} shows the performance of AMFT algorithm compared to Spark. With the absence of a failure, AMFT
algorithm outperforms spark FP-Growth version with an average speed-up of 20x with $\theta=0.01$ and an average speed-up of 8.6x
with $\theta=0.03$. The average speed-up in the case of smaller threshold ($\theta=0.01$) is larger because the size of
checkpointed FP-Trees is larger. Moreover, when checkpointing, the scalability of AMFT algorithm is better than the Spark-based algorithm
because AMFT only depends on checkpointing FP-Trees  and a set of transactions periodically,   which are both small with larger
number of cores. However, Spark depends on the RDD mechanism by having in-memory replication of both FP-Trees and transactions,
 overhead of  which increases  with a larger number of cores.

In the case of  a failure, the average gained speed-up from using AMFT compared to Spark is 15.3x with $\theta=0.01$ and 8.34x
with $\theta=0.03$. Performance of both AMFT and Spark-based algorithms becomes better with larger number of cores and/or smaller
support threshold (i.e., $\theta=0.03)$ because recovered FP-Tree is smaller in both cases.

\section{Related Work}
\label{sec:related}
Several researchers have proposed  FP-Growth algorithms
for both single node  and distributed memory systems~\cite{li2008pfp,modgi2014mining,bhokare2014novel,zhou2010balanced,buehrer2007toward,pramudiono2003parallel}.
These algorithms have addressed several issues for scalable FP-Growth such as memory
utilization, communication cost, and load-balancing. However, fault tolerance has not
been considered in these  efforts.

Several programming models proposed recently provide
automatic fault tolerance using functional paradigms. These include
MapReduce implementations like  Hadoop and Spark, as well as  MillWheel.
There have been studies for using
MapReduce  to parallelize frequent pattern mining algorithms,
 including   FP-Growth~\cite{li2008pfp,zhou2010balanced,itkar2013distributed}
and apriori~\cite{modgi2014mining,bhokare2014novel}. In these work,
MapReduce achieves fault-tolerance by re-executing all the tasks of the
failed node(s). As far as we are aware,
recovery algorithm has to completely re-execute the FP-Tree generation
from scratch in these implementations,  which severely and negatively impacts the recovery performance.

Scalable Checkpoint/Restart library (SCR) is another way  to support fault tolerant MPI-based applications through a multi-level
checkpointing technique~\cite{moody2010design}. SCR handles hardware failures in MPI application by performing less frequent and
inexpensive checkpoints  on available compute nodes memory. Our work has somewhat similar ideas, but further specializes them
by considering algorithm-specific properties.


\section{Acknowledgement}
The research described in this paper is
part of the Analysis in Motion (AIM) Laboratory Directed Research and
Development (LDRD) Initiative at Pacific Northwest National
Laboratory. 

\section{Conclusion}
\label{sec:summary}

This paper focuses on building a fault tolerance framework to support
FP-Growth algorithm in parallel systems. Three fault tolerance
algorithms have been proposed: Disk-based Fault Tolerance (DFT),
Synchronous Memory-based  Fault Tolerance  (SMFT), Asynchronous
Memory-based Fault Tolerance (AMFT). DFT algorithm represents the
brute-force approach to build a fault tolerance system using
periodically checkpoints on disk. However, the other two algorithms,
i.e., SMFT and AMFT, perform periodically checkpoints on memory instead
of disk to avoid I/O latency.

In SMFT algorithm, we shrink the processed transactions space and allocate a new space that can remotely be  accessed by other processes to perform FP-Tree and transactions checkpoint. This algorithm requires synchronization between processes before any single checkpoint  which adds more overhead to checkpointing operation. However, in AMFT algorithm, we use the transactions vector itself as checkpoint space to avoid any communication between processes during the checkpointing operation.


An extensive evaluation over 256, 512, 1024, and 2048 cores has been
performed on large datasets, i.e., 100 and 200 million transactions
datasets. Our evaluation demonstrates excellent efficiency for
checkpointing and recovery in comparison to the disk-based algorithm. Our
detailed experimental evaluation also shows low overheads and how we can
outperform Spark by an average of 20x with $\theta=0.01$ and 8.6x with $\theta=0.03$.

\bibliographystyle{abbrv}
\bibliography{vishnu}

\begin{thebibliography}{10}

\bibitem{abdulah2016addressing}
S.~M. S.~A. Abdulah.
\newblock {\em Addressing Disk Bandwidth Wall and Fault-Tolerance for
  Data-intensive Applications}.
\newblock PhD thesis, The Ohio State University, 2016.

\bibitem{agrawal2009quest}
R.~Agrawal and R.~Srikant.
\newblock Quest synthetic data generator. ibm almaden research center, san
  jose, california, 2009.

\bibitem{millwheel}
T.~Akidau, A.~Balikov, K.~Bekiro\u{g}lu, S.~Chernyak, J.~Haberman, R.~Lax,
  S.~McVeety, D.~Mills, P.~Nordstrom, and S.~Whittle.
\newblock Millwheel: Fault-tolerant stream processing at internet scale.
\newblock {\em Proc. VLDB Endow.}, 6(11):1033--1044, Aug. 2013.

\bibitem{alexandrov:95}
A.~Alexandrov, M.~F. Ionescu, K.~E. Schauser, and C.~Scheiman.
\newblock Loggp: incorporating long messages into the logp model—one step
  closer towards a realistic model for parallel computation.
\newblock In {\em Proceedings of the seventh annual ACM symposium on Parallel
  algorithms and architectures}, pages 95--105. ACM, 1995.

\bibitem{besta:hpdc14}
M.~Besta and T.~Hoefler.
\newblock Fault tolerance for remote memory access programming models.
\newblock In {\em Proceedings of the 23rd international symposium on
  High-performance parallel and distributed computing}, pages 37--48. ACM,
  2014.

\bibitem{bhokare2014novel}
P.~S. Bhokare and R.~Sharma.
\newblock A novel algorithm pda (parallel and distributed apriori) for frequent
  pattern mining.
\newblock {\em International Journal of Engineering}, 3(8), 2014.

\bibitem{bronevetsky:ics08}
G.~Bronevetsky and B.~de~Supinski.
\newblock Soft error vulnerability of iterative linear algebra methods.
\newblock In {\em Proceedings of the 22Nd Annual International Conference on
  Supercomputing}, ICS '08, 2008.

\bibitem{bronevetsky2003automated}
G.~Bronevetsky, D.~Marques, K.~Pingali, and P.~Stodghill.
\newblock Automated application-level checkpointing of mpi programs.
\newblock {\em ACM Sigplan Notices}, 38(10):84--94, 2003.

\bibitem{buehrer2006out}
G.~Buehrer, S.~Parthasarathy, and A.~Ghoting.
\newblock Out-of-core frequent pattern mining on a commodity pc.
\newblock In {\em Proceedings of the 12th ACM SIGKDD international conference
  on Knowledge discovery and data mining}, pages 86--95. ACM, 2006.

\bibitem{buehrer2007toward}
G.~Buehrer, S.~Parthasarathy, S.~Tatikonda, T.~Kurc, and J.~Saltz.
\newblock Toward terabyte pattern mining: an architecture-conscious solution.
\newblock In {\em Proceedings of the 12th ACM SIGPLAN symposium on Principles
  and practice of parallel programming}, pages 2--12. ACM, 2007.

\bibitem{buehrer:ppopp07}
G.~Buehrer, S.~Parthasarathy, S.~Tatikonda, T.~Kurc, and J.~Saltz.
\newblock Toward terabyte pattern mining: an architecture-conscious solution.
\newblock In {\em Proceedings of the 12th ACM SIGPLAN symposium on Principles
  and practice of parallel programming}, PPoPP '07, pages 2--12, 2007.

\bibitem{chen2011algorithm}
Z.~Chen.
\newblock Algorithm-based recovery for iterative methods without checkpointing.
\newblock In {\em Proceedings of the 20th international symposium on High
  performance distributed computing}, pages 73--84. ACM, 2011.

\bibitem{fang:damon09}
W.~Fang, M.~Lu, X.~Xiao, B.~He, and Q.~Luo.
\newblock Frequent itemset mining on graphics processors.
\newblock In {\em Proceedings of the Fifth International Workshop on Data
  Management on New Hardware}, DaMoN '09, pages 34--42, 2009.

\bibitem{mpi2}
A.~Geist, W.~Gropp, S.~Huss-Lederman, A.~Lumsdaine, E.~L. Lusk, W.~Saphir,
  T.~Skjellum, and M.~Snir.
\newblock {MPI}-2: Extending the message-passing interface.
\newblock In {\em Euro-Par, Vol. I}, pages 128--135, 1996.

\bibitem{gouda2005genmax}
K.~Gouda and M.~J. Zaki.
\newblock Genmax: An efficient algorithm for mining maximal frequent itemsets.
\newblock {\em Data Mining and Knowledge Discovery}, 11(3):223--242, 2005.

\bibitem{mpi1}
W.~Gropp, E.~Lusk, N.~Doss, and A.~Skjellum.
\newblock {A High-Performance, Portable Implementation of the {MPI} Message
  Passing Interface Standard}.
\newblock {\em Parallel Computing}, 22(6):789--828, 1996.

\bibitem{han2000mining}
J.~Han, J.~Pei, and Y.~Yin.
\newblock Mining frequent patterns without candidate generation.
\newblock In {\em ACM SIGMOD Record}, volume~29, pages 1--12. ACM, 2000.

\bibitem{hoefler2009towards}
T.~Hoefler, A.~Lumsdaine, and J.~Dongarra.
\newblock Towards efficient mapreduce using mpi.
\newblock In {\em Recent Advances in Parallel Virtual Machine and Message
  Passing Interface}, pages 240--249. Springer, 2009.

\bibitem{itkar2013distributed}
S.~A. Itkar and U.~V. Kulkarni.
\newblock Distributed algorithm for frequent pattern mining using hadoopmap
  reduce framework.
\newblock {\em International Conference on Advances in Computer Science,
  AETACS}, 2013.

\bibitem{li:recsys08}
H.~Li, Y.~Wang, D.~Zhang, M.~Zhang, and E.~Y. Chang.
\newblock Pfp: parallel fp-growth for query recommendation.
\newblock In {\em Proceedings of the 2008 ACM conference on Recommender
  systems}, RecSys '08, pages 107--114, 2008.

\bibitem{li2008pfp}
H.~Li, Y.~Wang, D.~Zhang, M.~Zhang, and E.~Y. Chang.
\newblock Pfp: parallel fp-growth for query recommendation.
\newblock In {\em Proceedings of the 2008 ACM conference on Recommender
  systems}, pages 107--114. ACM, 2008.

\bibitem{lin2014determining}
W.-T. Lin and C.-P. Chu.
\newblock Determining the appropriate number of nodes for fast mining of
  frequent patterns in distributed computing environments.
\newblock {\em International Journal of Parallel, Emergent and Distributed
  Systems}, (ahead-of-print):1--13, 2014.

\bibitem{lisboa2008algorithm}
C.~Lisboa, C.~Argyrides, D.~K. Pradhan, and L.~Carro.
\newblock Algorithm level fault tolerance: a technique to cope with long
  duration transient faults in matrix multiplication algorithms.
\newblock In {\em VLSI Test Symposium, 2008. VTS 2008. 26th IEEE}, pages
  363--370. IEEE, 2008.

\bibitem{modgi2014mining}
M.~P. Modgi and D.~Vaghela.
\newblock Mining distributed frequent itemset with hadoop.
\newblock {\em International Journal of Computer Science \& Information
  Technologies}, 5(3), 2014.

\bibitem{moody2010design}
A.~Moody, G.~Bronevetsky, K.~Mohror, and B.~R. De~Supinski.
\newblock Design, modeling, and evaluation of a scalable multi-level
  checkpointing system.
\newblock In {\em High Performance Computing, Networking, Storage and Analysis
  (SC), 2010 International Conference for}, pages 1--11. IEEE, 2010.

\bibitem{fpgrowth:cluster}
I.~Pramudiono and M.~Kitsuregawa.
\newblock Parallel fp-growth on pc cluster.
\newblock In {\em Proceedings of the 7th Pacific-Asia conference on Advances in
  knowledge discovery and data mining}, PAKDD'03, pages 467--473, 2003.

\bibitem{pramudiono2003parallel}
I.~Pramudiono and M.~Kitsuregawa.
\newblock Parallel fp-growth on pc cluster.
\newblock In {\em Advances in Knowledge Discovery and Data Mining}, pages
  467--473. Springer, 2003.

\bibitem{saini2014affinity}
A.~Saini, A.~Rezaei, F.~Mueller, P.~Hargrove, and E.~Roman.
\newblock {\em Affinity-aware checkpoint restart}.
\newblock North Carolina State University, 2014.

\bibitem{schroeder:dsn06}
B.~Schroeder and G.~A. Gibson.
\newblock A large-scale study of failures in high-performance computing
  systems.
\newblock In {\em Proceedings of the International Conference on Dependable
  Systems and Networks}, DSN '06, pages 249--258, Washington, DC, USA, 2006.
  IEEE Computer Society.

\bibitem{shantharam:ics11}
M.~Shantharam, S.~Srinivasmurthy, and P.~Raghavan.
\newblock Characterizing the impact of soft errors on iterative methods in
  scientific computing.
\newblock In {\em Proceedings of the International Conference on
  Supercomputing}, ICS '11, 2011.

\bibitem{shohdy2016parallel}
S.~Shohdy, A.~Vishnu, and G.~Agrawal.
\newblock Fault tolerant support vector machines.
\newblock In {\em Parallel Processing (ICPP), 2016 45th International
  Conference on}. IEEE, 2016.

\bibitem{sridharan:sc12}
V.~Sridharan and D.~Liberty.
\newblock A study of dram failures in the field.
\newblock In {\em Proceedings of the International Conference on High
  Performance Computing, Networking, Storage and Analysis}, SC '12, pages
  76:1--76:11, 2012.

\bibitem{sridharan:sc13}
V.~Sridharan, J.~Stearley, N.~DeBardeleben, S.~Blanchard, and S.~Gurumurthi.
\newblock Feng shui of supercomputer memory: Positional effects in dram and
  sram faults.
\newblock In {\em Proceedings of the International Conference on High
  Performance Computing, Networking, Storage and Analysis}, SC '13, pages
  22:1--22:11, New York, NY, USA, 2013. ACM.

\bibitem{vandam:jctc13}
H.~J.~J. van Dam, A.~Vishnu, and W.~A. de~Jong.
\newblock A case for soft error detection and correction in computational
  chemistry.
\newblock {\em Journal of Chemical Theory and Computation}, 9(9), 2013.

\bibitem{vishnu:cluster15}
A.~Vishnu and K.~Agarwal.
\newblock Ieee cluster.
\newblock chapter Large Scale Frequent Pattern Mining using MPI One-sided
  Model. 2015.

\bibitem{xu2014efficient}
J.~Xu, N.~Li, X.-J. Mao, and Y.-B. Yang.
\newblock Efficient probabilistic frequent itemset mining in big sparse
  uncertain data.
\newblock In {\em PRICAI 2014: Trends in Artificial Intelligence}, pages
  235--247. Springer, 2014.

\bibitem{yu2011efficient}
K.-M. Yu and S.-H. Wu.
\newblock An efficient load balancing multi-core frequent patterns mining
  algorithm.
\newblock In {\em Trust, Security and Privacy in Computing and Communications
  (TrustCom), 2011 IEEE 10th International Conference on}, pages 1408--1412.
  IEEE, 2011.

\bibitem{spark}
M.~Zaharia, M.~Chowdhury, T.~Das, A.~Dave, J.~Ma, M.~McCauley, M.~J. Franklin,
  S.~Shenker, and I.~Stoica.
\newblock Resilient distributed datasets: A fault-tolerant abstraction for
  in-memory cluster computing.
\newblock In {\em Proceedings of the 9th USENIX Conference on Networked Systems
  Design and Implementation}, NSDI'12, pages 2--2, Berkeley, CA, USA, 2012.
  USENIX Association.

\bibitem{zaki1997parallel}
M.~J. Zaki, S.~Parthasarathy, M.~Ogihara, and W.~Li.
\newblock Parallel algorithms for discovery of association rules.
\newblock {\em Data Mining and Knowledge Discovery}, 1(4):343--373, 1997.

\bibitem{zhou2010balanced}
L.~Zhou, Z.~Zhong, J.~Chang, J.~Li, J.~Huang, and S.~Feng.
\newblock Balanced parallel fp-growth with mapreduce.
\newblock In {\em Information Computing and Telecommunications (YC-ICT), 2010
  IEEE Youth Conference on}, pages 243--246. IEEE, 2010.

\end{thebibliography}

\end{document}